\documentclass{gGAF2e}

\RequirePackage{hyperref}
\hypersetup{
	colorlinks = true,
	linkcolor = blue,
	citecolor = blue,
	filecolor = blue,
	urlcolor = blue,
	pdfborder = {0 0 0}
}

\usepackage{color}
\usepackage{bm}

\renewcommand{\vec}[1]{\bm{#1}}

\graphicspath{{./}{./figures/}}

\newcommand{\apos}[1]{{\char13}#1{\char13}}
\newcommand{\tab}[1]{table\,\ref{T:#1}}
\newcommand{\sect}[1]{section\,\ref{S:#1}}

\newcommand{\fig}[1]{figure\,\ref{F:#1}}
\newcommand{\figs}[2]{figures\,\ref{F:#1} and \ref{F:#2}}

\newcommand{\eqn}[1]{equation\,(\ref{E:#1})}

\newcommand{\graphflex}[4][figure]{\begin{#1}#2\caption{#4\label{F:#3}}\end{#1}}

\newcommand{\graphwidthflex}[6][figure*]{\graphflex[#1]{#5\includegraphics[width=#4]{#2.pdf}}{#3}{#6}}
\newcommand{\graphwidth}[4][12cm]{\graphwidthflex{#2}{#3}{#1}{\centering}{#4}}
\newcommand{\graphfull}[3]{\graphwidth[\textwidth]{#1}{#2}{#3}}

\newcommand{\graphflexbottom}[4][figure]{\begin{#1}[b]#2\caption{#4\label{F:#3}}\end{#1}}
\newcommand{\graphwidthflexbottom}[6][figure*]{\graphflexbottom[#1]{#5\includegraphics[width=#4]{#2.pdf}}{#3}{#6}}
\newcommand{\graphwidthbottom}[4][12cm]{\graphwidthflexbottom{#2}{#3}{#1}{\centering}{#4}}
\newcommand{\graphfullbottom}[3]{\graphwidthbottom[\textwidth]{#1}{#2}{#3}}
\newcommand{\eql}[1]{\begin{equation}#1\end{equation}}
\newcommand{\eqa}[1]{\begin{eqnarray}#1\end{eqnarray}}

\newcommand{\eqi}[1]{\hspace{0.4mm}$#1$}

\DeclareRobustCommand*{\unit}[1]{\def~{\,}\ensuremath{\mathrm{\,#1}}}
\definecolor{darkgreen}{rgb}{0.1,0.6,0.1}
\newcommand{\red}[1]{{{\color{red}#1}}}

\begin{document}

\jvol{00} \jnum{00} \jyear{2019} \jmonth{July}
\doi{10.1080/03091929.2019.1643849}

\markboth{Ph.-A.~Bourdin}{Geophysical \& Astrophysical Fluid Dynamics}


\title{Driving solar coronal MHD simulations on high-performance computers}

\author{Philippe-A.~Bourdin}
\email{Philippe.Bourdin@oeaw.ac.at}
\orcid{0000-0002-6793-601X}
\affiliation{Space Research Institute, Austrian Academy of Sciences, Graz, Austria}

\received{30 August 2018}
\accepted{11 July 2019}

\maketitle

\begin{abstract}
The quality of today's research is often tightly limited to the available computing power and scalability of codes to many processors.
For example, tackling the problem of heating the solar corona requires a most realistic description of the plasma dynamics and the magnetic field.
Numerically solving such a magneto-hydrodynamical (MHD) description of a small active region (AR) on the Sun requires millions of computation hours on current high-performance computing (HPC) hardware.
The aim of this work is to describe methods for an efficient parallelization of boundary conditions and data input/output (IO) strategies that allow for a better scaling towards thousands of processors (CPUs).
The {\em Pencil Code} is tested before and after optimization to compare the performance and scalability of a coronal MHD model above an AR.
We present a novel boundary condition for non-vertical magnetic fields in the photosphere, where we approach the realistic pressure increase below the photosphere.
With that, magnetic flux bundles become narrower with depth and the flux density increases accordingly.
The scalability is improved by more than one order of magnitude through the HPC-friendly boundary conditions and IO strategies.
This work describes also the necessary nudging methods to drive the MHD model with observed magnetic fields from the Sun's photosphere.
In addition, we present the upper and lower atmospheric boundary conditions (photospheric and towards the outer corona), including swamp layers to diminish perturbations before they reach the boundaries.
Altogether, these methods enable more realistic 3D~MHD simulations than previous models regarding the coronal heating problem above an AR -- simply because of the ability to use a large amount of CPUs efficiently in parallel.
\end{abstract}

\section{Introduction} \label{S:introduction}

Cutting-edge science is often limited only by computational resources.
More realistic models come into reach with the continuously increasing computer power.
One such fundamental step towards better understanding the famous coronal heating problem and plasma heating mechanisms was achieved with the work of \cite{Bourdin+al:2013_overview} and their following publications.
This step only became feasible, because the {\em{Pencil Code}} got extended with some massive-parallel methods that enabled large-scale models to be run within the time of a typical PhD contract.
Otherwise, scaling to less processors, the main simulation run described in \cite{Bourdin+al:2013_overview} would have taken about four years using \eqi{256} processors instead of about one year it took with \eqi{1024} processors.

The coronal heating mechanisms are unclear since many decades \citep{Klimchuk:2006}.
For a better understanding of the coronal heating, novel models need to be as realistic as possible, so that the most relevant physical processes may be captured and analyzed in a data post-processing step.
It is currently not possible to use only observations for that task, because some key quantities like the coronal 3D structure and amplitude of the magnetic field still remain inaccessible.
Still, long-standing theories need to be verified, like the entangling of magnetic field lines in the corona through shuffling of footpoints in the photosphere \citep{Parker:1972}.
This could lead to so-called nanoflares that heat the corona through small and short-lived magnetic reconnection events after the field became entangled in the corona \citep{Parker:1988}.
Magneto-hydrodynamic (MHD) turbulence and Alfv{\'e}n wave turbulence compete to explain the coronal heat input; see the results of \cite{Rappazzo+al:2007,Rappazzo+al:2008} versus results from \cite{vanBallegooijen:2011,vanBallegooijen:2014}.
More recent models may test these scenarios against a realistic coronal magnetic field configuration \citep{Bourdin+al:2016_scaling-laws}.

We complement the approach of \cite{Rempel:2012} that includes parts of the convection zone below the photosphere to drive the magnetic field.
While \cite{Rempel:2017} uses more advanced physics, our models are driven by actual photospheric observations and hence are able to match observed structures in the corona directly.
This work is a continuation of {\em{Pencil Code}} models, like first presented in \cite{Bingert+al:2010} that led to a better understanding on why the cross-section of coronal loops appears as roughly constant \citep{Peter+Bingert:2012}.
Still, we do not know if such loops have a single- or multi-stranded structure in their magnetic fields \citep{Peter+al:2013}.
\cite{Chen+al:2014} use a convection simulation with two emerging sunspots to drive a separate corona model.
\cite{Peter+al:2015} review what MHD models can tell us on coronal heating mechanisms.
For different phenomena, like type-II spicules or a more realistic chromosphere, other groups include radiative transfer or ambipolar diffusion in their equation sets \citep{Hansteen+al:2010,Martinez-Sykora+al:2011,Wedemeyer-Böhm+al:2012}.

We note that the Alfv{\'e}n velocity is at least about one order of magnitude larger in the corona than in the chromosphere, which suggests that any changes in the chromospheric magnetic field would almost instantly and quasi-statically change the coronal field configuration \citep{Bourdin+al:2014_coronal-loops,Bourdin+al:2015_energy-input}.
The field may either reconnect and release magnetic twist into the solar wind, the twisting magnetic perturbation may leave the corona before a substantial twist is build up on ``open'' field lines that connect to the heliosphere, or the perturbation may cross the whole corona on closed loops and eventually twist the chromospheric field on the other end of the loop, where the Alfv{\'e}n velocity becomes lower again.
To address this topic, we need to know the coronal field more precisely and we may indeed track coronal field lines from 3D~MHD model data \citep{Bourdin+al:2018_helicity}.

In order to obtain realistic coronal magnetic fields, we need to drive the MHD simulation with observations of real solar magnetograms, e.g. of a full active region (AR) that features some coronal EUV-bright loops that are known to be at least 1 \unit{MK} hot.
We cover an AR with some surrounding quiet Sun (about \eqi{240 \times 240\unit{Mm^2}}) with \eqi{1024 \times 1024} grid points.
The vertical grid resolution varies form about 100\unit{km} for the photosphere and chromosphere to about 800\unit{km} in the upper corona.
Furthermore, the photospheric horizontal shifting motions from granulation are required to ultimately test the field-line braiding mechanism.
To this end, we implemented a photospheric granulation driver in the \texttt{solar\_corona} module of the {\em{Pencil Code}} together with schemes for photospheric and chromospheric nudging that always gently push the model towards the observed state in the photosphere and that provide a lower solar atmosphere that adapts to pressure and temperature stratification changes in the corona.
Our chromosphere acts here as a flexible lower boundary condition and as a reservoir for mass and internal energy, which is also the case in reality, because the lower atmosphere hosts significantly more mass than the corona.

With a novel boundary condition for the photospheric magnetic field we allow the granulation driver to push the foot points of field lines.
This changes the horizontal component of the magnetic field already in the photosphere and in the grid cells above, which is required to test the braiding mechanism proposed by \cite{Parker:1972}.
Therefore, we need to extrapolate the magnetic field to the interior of the Sun, in order to provide the required ghost zones for the computation of numerical dervatives.
Because the pressure increases below the photosphere, magnetic flux tubes shrink in their diameter with depth; see \sect{inverted.extrapolation} for details.

In the following we describe the photospheric and chromospheric nudging, the granulation driver, a new magnetic-field extrapolation, as well as coronal boundary conditions.
Finally, we analyze the parallelization of the Fourier transform and the data access (IO) routines.

\section{Photospheric nudging}

The photosphere is the lower boundary condition for our simulation domain.
This requires us to set the temperature \eqi{T}, the density \eqi{\rho}, and the vertical components of the vector quantities that are perpendicular to this boundary, the vertical velocity \eqi{u_z} and the magnetic field \eqi{B_z}.
In the photosphere, the convective cells from below practically reach the layer, where radiation efficiently cools the plasma.
Hence, the advective heat transport by convection ends and the vertical transport of hot plasma breaks into granules, where the plasma motion has a horizontal component  of some \unit{km/s} at about \eqi{0.5-0.8\unit{Mm}} around the granule center \citep{RuizCobo+al:1996}.
Once cooled, the now denser plasma gets submerged below the surface due to gravity and will eventually be heated again to complete the convection cycle.

In the same time, we know that plasma beta, the ratio between thermal and magnetic pressure, is near or above unity at the photosphere, in average \citep{Bourdin:2017_beta}.
As a result, the horizontal motions of breaking granules will advect the magnetic field with them.
Many subsequently rising and decaying granules may push the field lines even on spatial scales larger than one single granule and in a random-walk fashion.
Finally, magnetic flux bundles are rooted below the photosphere and constantly undergo a smooth global reconfiguration due to dynamo processes, which leads to horizontal motions of these patches on spatial scales much larger than one granule, but typically with a significantly lower speed.
To complete the spectrum of photospheric driving motions, we will need to consistently combine both, small-scale granular motions and the large-scale magnetic reconfiguration in the photosphere.

\subsection{Magnetic field}

\graphfullbottom{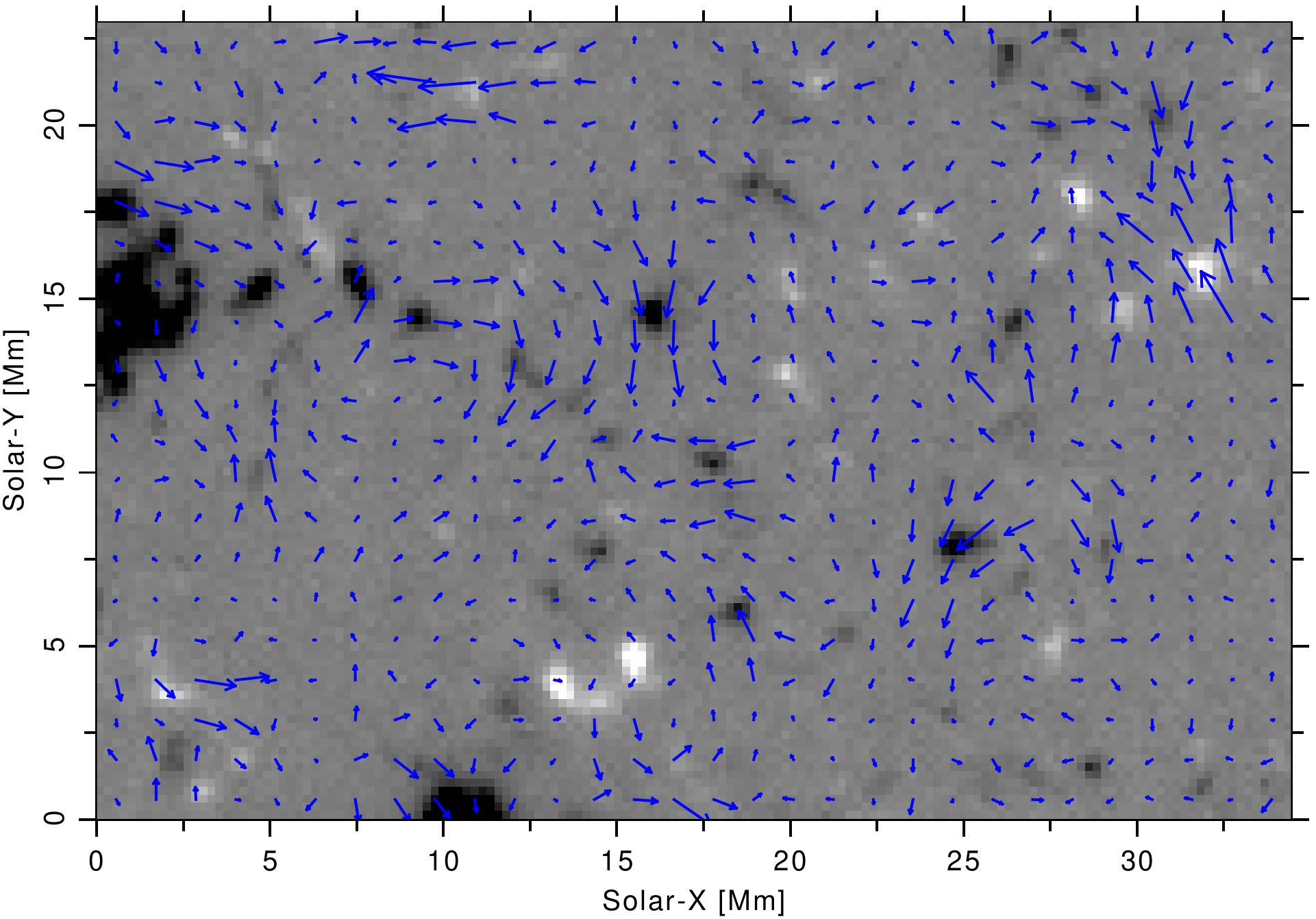}{magnetogram-LCT}{Hinode line-of-sight magnetogram (grayscale, saturated at \eqi{\pm300\unit{G}}) with overlaid velocity vectors (blue) obtained with a local correlation tracking. The observation was made near disc center and is from 2007~November~14.}

For the photospheric magnetic field nudging we use actual observations of the {\em Hinode} satellite, where the data from {\em{SOT/NFI}} (Solar Optical Telescope/Narrowband Filter Imager) give us the line-of-sight (LOS) component of the magnetic field in the photosphere \citep{Kosugi+al:2007,Tsuneta+al:2008}; see gray-scale background in \fig{magnetogram-LCT}.
The resolution of our LOS magnetograms is about \eqi{120\unit{km}} per pixel.
Magnetograms do not resemble visible-light intensity observations \citep[cf.][]{Bourdin:2011} and hence we do neither directly see the granules in such magnetograms nor could we derive the horizontal plasma motions.
Even intensity maps of the photosphere from the same telescope would not resolve a granule with more than a few pixels.
Therefore, we can not recover the horizontal granular advection from {\em Hinode} data and have to find another way to mimic realistic granulation; see \sect{gran.driver}.

The LOS magnetograms are uncalibrated and we use the smaller field-of-view (FOV) of the {\em{SOT/SP}} (Spectro-Polarimetric) instrument \citep{Lites+al:2013} that provides calibrated vector magnetograms and co-align both, the LOS and {\em{SP}} data, while we reduce the resolution of the LOS to the {\em{SP}} data.
A pixel by pixel comparison allows us to find the calibration factor for the full-FOV LOS magnetograms.
For consitency, we also interpolate two LOS magnetograms in time to match the exact {\em{SP}} observation time.
Because both instrument's detectors have a slight rotation against each other, we split the common FOV in about \eqi{8 \times 8} subfields and coaling them separately.
These steps help to reduce the broadness of the scatter in the distribution of the correlated data that we show in \fig{woehl-plot}.

The horizontal extent of our {\em{NFI}} FOV is about one seventh of the solar radius.
We correct the LOS magnetograms for the surface curvature within the FOV of {\em{SOT/NFI}} by assuming that all strong flux concentrations are mainly vertical, so that we may scale the magnetic field amplitude of both polarities with \eqi{1/\cos\theta}, where \eqi{\theta} is the angle between the LOS and the vertical direction.
We now co-align the time series of magnetograms to the one in middle of the time series so that we correct for spacecraft jitter and the rotating Sun.
Finally, we crop all images to their common FOV.

\graphwidth[0.99\textwidth]{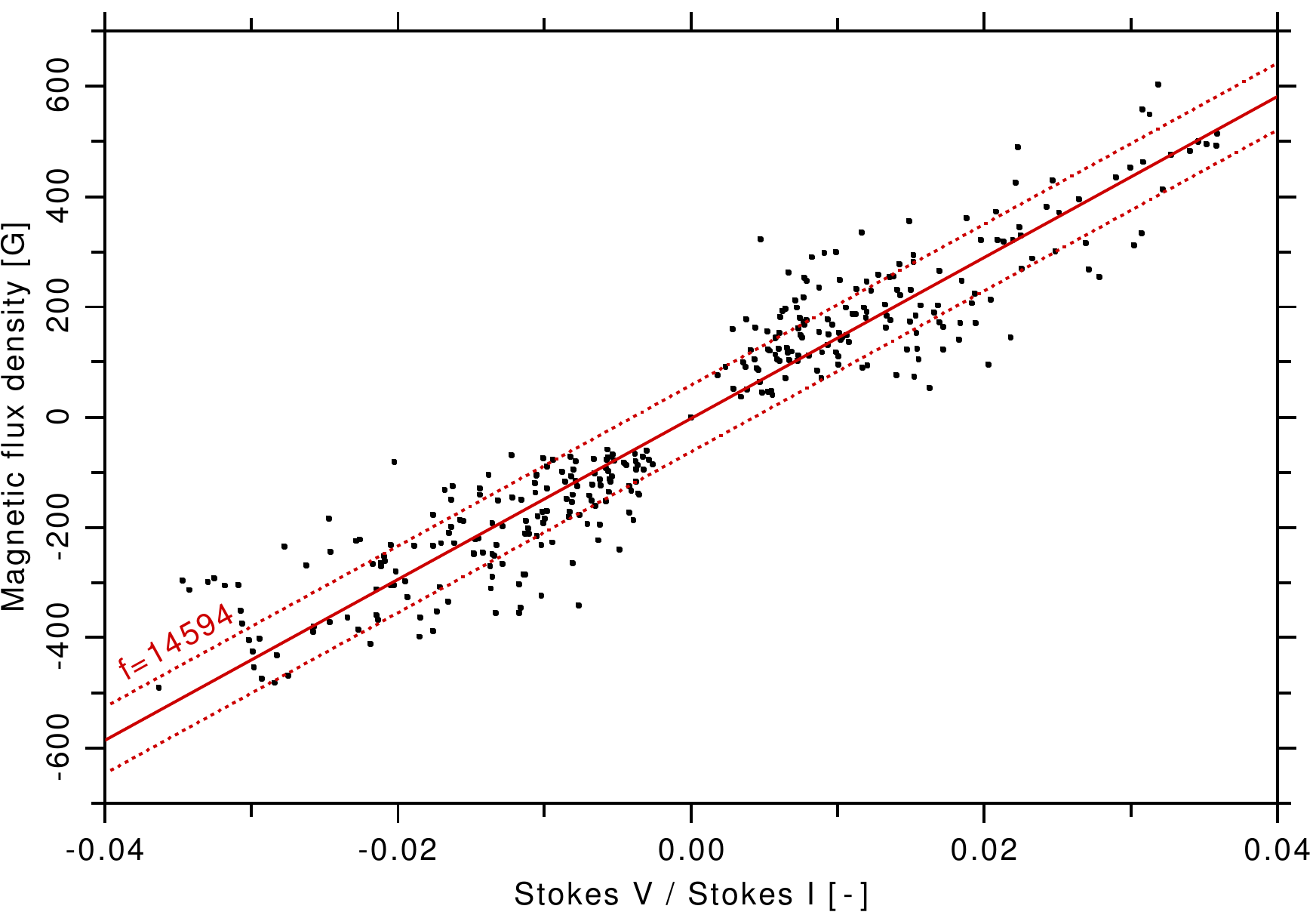}{woehl-plot}{Distribution of co-aligned uncalibrated {\em{SOT/NFI}} and calibrated {\em{SOT/SP}} data. The red line represents a least-absolute-deviation (LAD) fit with its uncertainty interval marked by dotted lines. Figure taken from \cite{Bourdin:2014:thesis}; see Figure~2.4 on page~31.}

For each of the magnetograms we rescale both polarities separately and with a constant factor to obtain flux-balanced observations.
This is required here to avoid an average magnetic monopole that would force all of its flux to leave through the upper boundary, because the simulation domain is periodic in the horizontal directions.
Also this avoids unwanted effects from a time-varying total flux that would have to propagate immediately through the whole domain because of our boundary conditions and hence disturb the actual coronal magnetic field in an unnatural way.
In any case, the total magnetic flux of the whole Sun is always zero.
Natural localized flux imbalances on the real Sun cause magnetic connectivity to remote regions, which we can not include in our model due to the limited simulation domain size.
The connectivity that we miss through our flux balancing is minor and does not significantly change the field geometry in our model.

When we use a non-periodic magnetogram for a periodic simulation domain, we need to overlap the boundaries of the magnetogram with a smooth transition at the borders of the observable FOV.
This makes the magnetograms periodic and we do not see artefacts in our model if the overlap region is about 24\ pixels or about 6\unit{Mm} wide.

If we would now simply impose the obtained flux-balanced time series of LOS magnetograms on our lower boundary for all times, the magnetic field description would on one hand be perfectly consistent with the observation.
On the other hand, we would effectively stop the granular driving to operate, because of two reasons:
First, the granular motions are not directly recovered with LOS magnetograms.
Second, our granulation driver can practically never advect a magnetic field line, because its original state would get restored in the next iteration of the simulation.
Therefore, we have to allow the granulation to advect the field lines and in the same time we have to carefully restore the observed magnetic state.
Luckily, small-scale granules live much shorter than the large-scale magnetic field needs to change substantially.
This allows us to apply two nudging processes on different time scales, where we push the model into the targeted states by an exponential decay with distinct characteristic half times for each process.

\subsection{Local correlation tracking}

Once we want to allow for time-varying magnetograms at the lower boundary, we need to apply also horizontal plasma motions that are consistent with the displacement of the observed magnetic patches.
When plasma beta is around unity, the magnetic field can be shifted from plasma motions or the plasma can be dragged together with moving field lines.
Both ways, the magnetic field displacement and the plasma motion should be consistent.
Therefore, we deduce the horizontal motions of magnetic patches with a local-correlation-tracking (LCT) method and obtain another lower-amplitude velocity field on spatial scales larger than granules.
This LCT velocity field we also need to apply in the photosphere.

Even though we omit here electric fields that are in principle necessary to fulfill the induction equation, the good match of the simulated AR corona with the observed one \citep{Bourdin+al:2013_overview} supports this simplification.
Still, the question remains how much the coronal heat input may change if we include those omitted photospheric electric fields.

In the first step to obtain a velocity field out of a magnetogram time series, we rebin the data to the optical resolution, which means we bin a square of \eqi{2 \times 2} pixels into one simulation pixel of about 235\unit{km} side length.
Then we compute the local cross-correlation coefficients of two consecutive frames of the time series, where we shift the frames to each other by one pixel in each direction.
Together with the original zero-shift correlation coefficient this gives us five coefficients, where we find the sub-pixel shift vector as the local maximum of a five-point Gaussian fit along two spatial dimensions.
We apply a Gaussian-convolution filter to the obtained map of shift vectors to remove small-scale fluctuations that we are not interested in.
Finally, we need to interpolate this vector field at the actual pixel positions of our simulation grid.

The resulting velocity field from our LCT method is displayed as blue arrows in \fig{magnetogram-LCT}.
As we see, some magnetic patches are surrounded by a local velocity field pointing in one direction, where the strongest velocity is near the center of that patch; see black patch at \eqi{x=25\unit{Mm}} and \eqi{y=8\unit{Mm}} in \fig{magnetogram-LCT}.
The correlation decays with distance from the flux concentrations, which is expected because one can not deduce a horizontal velocity from two consecutive insignificant flux regions (gray background).

\subsection{Granulation driver}\label{S:gran.driver}

\graphfullbottom{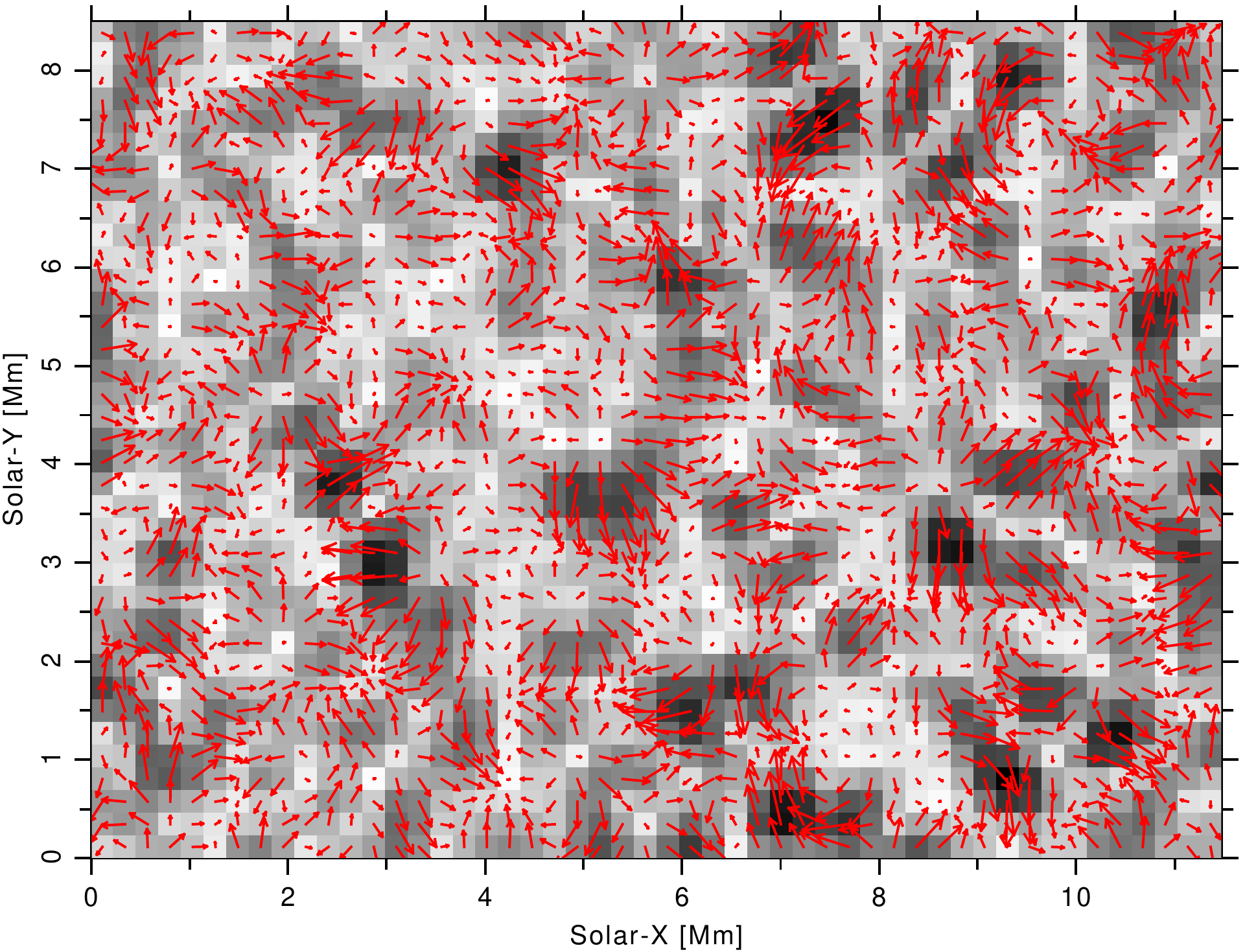}{granulation_driver_vectors}{Granulation driver velocity field (grayscale, saturated black is 8\unit{km/s}) with overplotted velocity vectors (red). One grayscale square represents one simulation pixel with a side length of 230\unit{km}.}

To mimic the horizontal advecting motions of breaking granules in the photosphere we use a constructed velocity field that resembles granules with a diameter of 1.6\unit{Mm}.
We modify here a method described by \cite{Gudiksen+Nordlund:2005a} that is based on a weighted {\em{Voronoi tessellation}} driver \citep{Schrijver+al:1997}.
In contrast to \cite{Gudiksen+Nordlund:2002} who use a random pattern for driving their simulation, we generate a velocity field that resembles solar granulation not only in a statistical sense.
Around each granule we define an additional radius of 0.24\unit{Mm} as a zone where inter-granular lanes may form due to overlapping of the velocity fields from neighboring granules.
The typical lifetime of a granule is \eqi{T_{\rm gran}=5\unit{min}}.
The amplitude of the typical velocity in the interior of a granule we define as \eqi{v_0=1.028\unit{km/s}}.
The radial velocity within each granule must of course be zero at the center, increases with distance from the center, until it decays outside of the granule's radius and within the inter-granular lane, which ranges from 0.8 to 1.04\unit{Mm} distance from the center.
An example of a mainly radial velocity field can be seen near \eqi{x=6} and \eqi{y=8\unit{Mm}} in \fig{granulation_driver_vectors}.

We randomly fill the simulation domain with granules, until there is no more available space for new granules.
No new granule is allowed to emerge within the inner radius of 0.64\unit{Mm} around any other granule's center.
A granule ceases to exist, when its velocity amplitude \eqi{v(t)} drops below the threshold of \eqi{v_{\rm min}=0.78 ~ v_0}; see \eqn{v_t}.
The former granule's area then becomes immediately available for new granules.

To avoid that all granules emerge and decay always at the same time, we vary the average granular lifetime with normally distributed random values within \eqi{\pm10\unit{\%}} and define this individual lifetime as
\eqi{T = T_{\rm gran}\pm10\unit{\%}}.
We also alter the typical granular velocity amplitude in the same way for each granule as \eqi{v_{\rm max} = v_0\pm15\unit{\%}}.
Accordingly, we change the time when the maximum velocity amplitude shall be reached (\eqi{t_{\rm max}}) to be earlier for shorter lifetimes and later for longer lifetimes \eqi{T} as
\eql{t_{\rm max} = T \sqrt{|\log ( v_{\rm min} / v_{\rm max} )|} .}

For a smooth emergence and decay of a granule, we multiply its velocity amplitude with a life-cycle function that depends on the simulation time \eqi{t} and the time when the granule was created \eqi{t_0}:
\eql{v(t) = v_{\rm max} \cdot \exp \left [ - \left ({t - t_0 - t_{\rm max}} \over T \right )^2 \right ] \label{E:v_t}}

Now that we have filled our FOV with granules, in particular with overlapping areas that have non-radial velocities, we may amplify the purely stochastic vorticity in our two-dimensional horizontal-velocity field \eqi{{\vec{v}}_{\rm gran}(\vec{r},t)} at the position vector \eqi{\vec{r}}.
For that, we Fourier-transform the velocity field, extract its rotational part \eqi{\hat{\vec{v}}_{\rm rot}}, and filter out large wavenumbers with an exponential weighting factor
\eql{{\hat{\vec{v}}_{\rm rot}}'(\vec{k},t) = \hat{\vec{v}}_{\rm rot}(\vec{k},t) \cdot \exp \left [ - \left (2 {|\vec{k}| / k_N} \right )^4 \right ],}
where \eqi{\vec{k}} is the in-plane wave vector and \eqi{k_N} is the Nyquist frequency.
We transform \eqi{{\hat{\vec{v}}_{\rm rot}}'(\vec{k},t)} back to real space \eqi{{\vec{v}}_{\rm rot}'(\vec{r},t)}, amplify it with a factor of \eqi{f_{\rm rot}}, and add it back to the initial granular field.
We choose \eqi{f_{\rm rot}=5} so that the resulting velocity field will have flows along the inter-granular lanes that reach realistic horizontal speeds of about \eqi{8\unit{km/s}}, which is similar to the observed inter-granular velocities; see saturated black color in \fig{granulation_driver_vectors}.

Finally, we correct the new velocity field with enhanced vorticity \eqi{{\vec{v}}_{\rm vor}} so that its root-mean-squared velocity becomes identical to one of the initial granulation field:
\eqa{{\vec{v}}_{\rm vor} &=& {\vec{v}}_{\rm gran} + f_{\rm rot} ~ {\vec{v}}_{\rm rot}' \nonumber \\
{\vec{v}}_{\rm gran}^\ast &=& {\vec{v}}_{\rm vor} \cdot \sqrt{\langle {{\vec{v}}_{\rm gran}}^2 \rangle / \langle {{\vec{v}}_{\rm vor}}^2 \rangle} }
We obtain a velocity field with radial outflows that turn into tangential velocities.
The tangential flows of neighboring granules may interfere and form inter-granular lanes with larger velocities than the average speeds inside the granules; see around \eqi{x=3} and \eqi{y=4\unit{Mm}} in \fig{granulation_driver_vectors} for an example.

For consistency with external velocity fields, like the LCT velocities derived from a magnetogram time series, all granule centers will be propagated in accordance with the pre-existing velocity vector at each granule's center.
If no other velocity field has been activated during the generation of the granular driving motions, the granule centers remain where they emerge.

A common problem in massive-parallel simulations are boundary conditions that involve only few processors (see \sect{FFT} for another example) because the boundary processors then cause inactive delays on the non-boundary processors.
The granulation driver code in the {\em{Pencil Code}} is not scalable.
Therefore, we improve the scalability of our coronal model with a two-step scheme:
First, we create the granulation in a 2D simulation of the photosphere and write the generated velocity field to an external file.
This step requires only few processors and may run with a substantially larger timestep than the full coronal model because we need only one granulation snapshot about every 10 seconds and some timestep-critical methods (like the Spitzer heat conduction) are not relevant here.
Second, in the 3D coronal model we read the previously generated velocity field, distribute this data to each of the photospheric boundary processors, and then interpolate between the granulation snapshots in time.
Because the timestep in the 3D model is substantially lower than 10 seconds, we save many iterations of the granulation driver code, because the changes therein are minimal.
In addition, the time-interpolation between two granulation snapshots can be scaled to all involved boundary processors, because each processor already has all necessary data.
Finally, a time-interpolation consists of much less computations than one granulation driver iteration.
This scheme reduces the waiting time of non-boundary processors to a minimum and hence makes the coronal model better scalable.%
\footnote{The related parameters in the \texttt{solar\_corona} module are the logical flag \texttt{lgranulation} to activate the granulation driver, \texttt{lwrite\_driver} to write the generated velocity field to an external file, \texttt{tau\_inv} as the inverse time scale for velodity-field nudging, \texttt{vorticity\_factor} as the factor to increase the vorticity, and \texttt{dt\_gran} as the update interval for the granules. The logical flags \texttt{luse\_vel\_field} and \texttt{luse\_mag\_vel\_field} activate the reading of external velocity fields.}


The photospheric velocity driver is of course not switched on immediately.
If we would do that, we would cause an instantaneous force acting on an equilibirum state.
This is equivalent to a very strong impulse that creates a shock front; see Figure~3 of \cite{Bourdin:2014_switch-on}.
To avoid this switch-on effect, we smoothly ramp up the targeted velocity field linearly within the first minute of the simulation.
In addition, the horizontal bulk velocity \eqi{\left. \vec{u}_{\rm hor} \right|_j} at the photospheric boundary and in the three ghost layers below (\eqi{j \in \{0,-1,-2,-3\}}) is never directly imposed.
We smoothly push the velocity to its target value by an exponential decay that we implement by an additional term in the equation of motion (shortened by `...'):
\eql{\left. \partial \vec{u}_{\rm hor} / \partial t \right|_j = ... -\tau_u \left (\left. \vec{u}_{\rm hor} \right|_j - {\vec{v}}_{\rm gran}^\ast \right )}
with the inverse decay half-time \eqi{\tau_u=0.5\unit{s^{-1}}}.

\fig{LCT_histogram} shows a direct comparison of LCT and granulation driver velocity histograms, where we see that the LCT velocities have a peak below 0.1\unit{km/s} and hence smaller amplitudes than the granulation driver that peaks around 2\unit{km/s}.
We checked that most of the driving power (Poynting flux) stems from the granulation driver.
The granular velocities vary on much smaller spatial scales than the LCT velocities from the horizontal movement of magnetic patches, as also becomes clear from a direct comparison of the axes in \figs{magnetogram-LCT}{granulation_driver_vectors}.

\graphfull{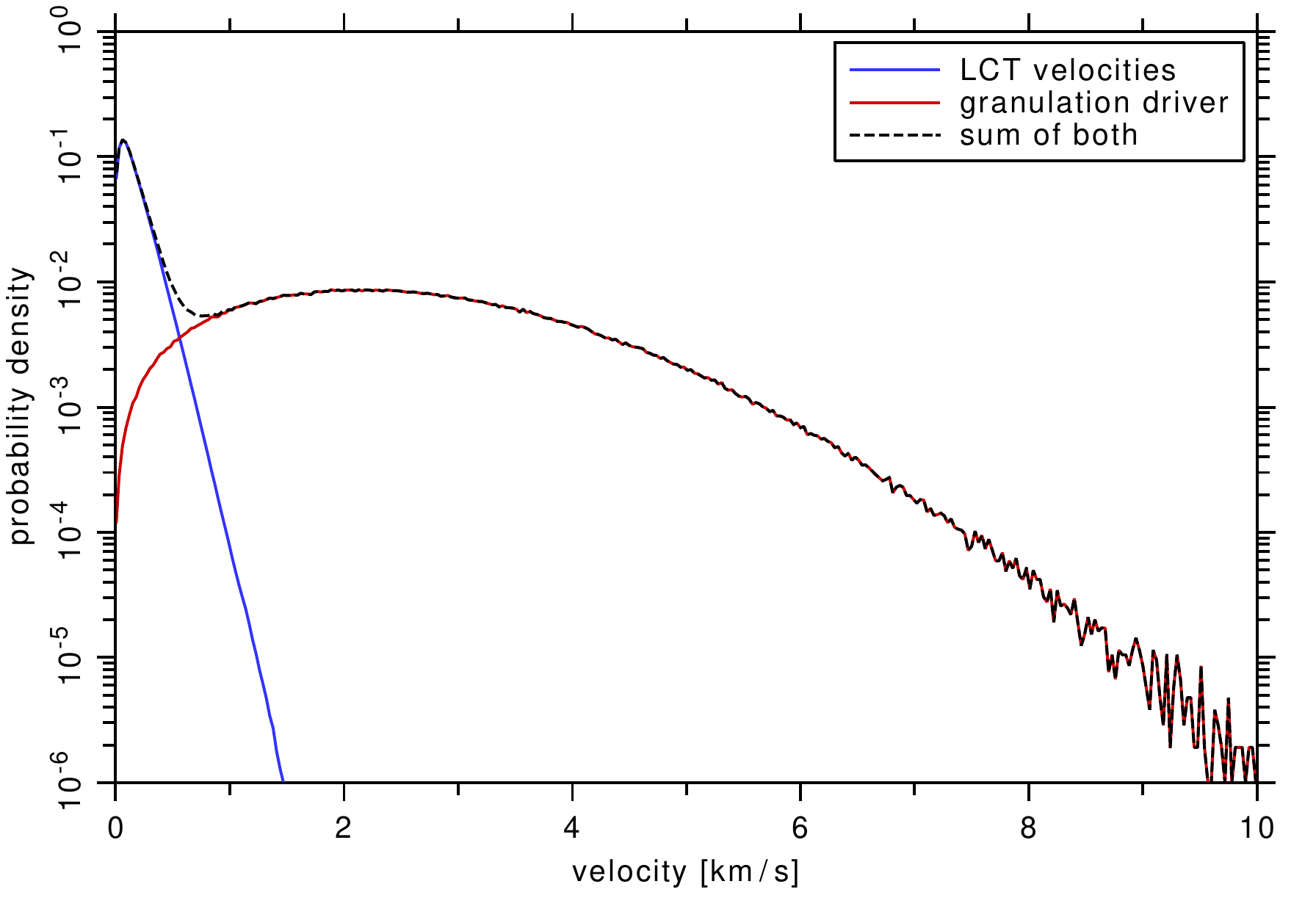}{LCT_histogram}{Histograms of LCT (blue) and granulation driver (red) velocities.}

\subsection{Magnetic-field extrapolation} \label{S:inverted.extrapolation}

We use the vector potential \eqi{\vec{A}} for our computation and the magnetic field \eqi{\vec{B}} is then only a derived quantity that is always divergence free and is unique for any given gauge \eqi{\Phi_0}:
\eql{\vec{B}=\vec{\nabla} \times (\vec{A} + \Phi_0)}
To prescribe the vertical magnetic field \eqi{B_z} in the photosphere, while using the vector potential \eqi{\vec{A}} within our simulation, we need to set the two components \eqi{A_x} and \eqi{A_y} because
\eql{B_z = {\partial A_y \over \partial x} - {\partial A_x \over \partial y} .}
\eqi{A_z} may then be set according to a self-chosen gauge, here the Weyl gauge with \eqi{\Phi_0 = 0}.

\graphfull{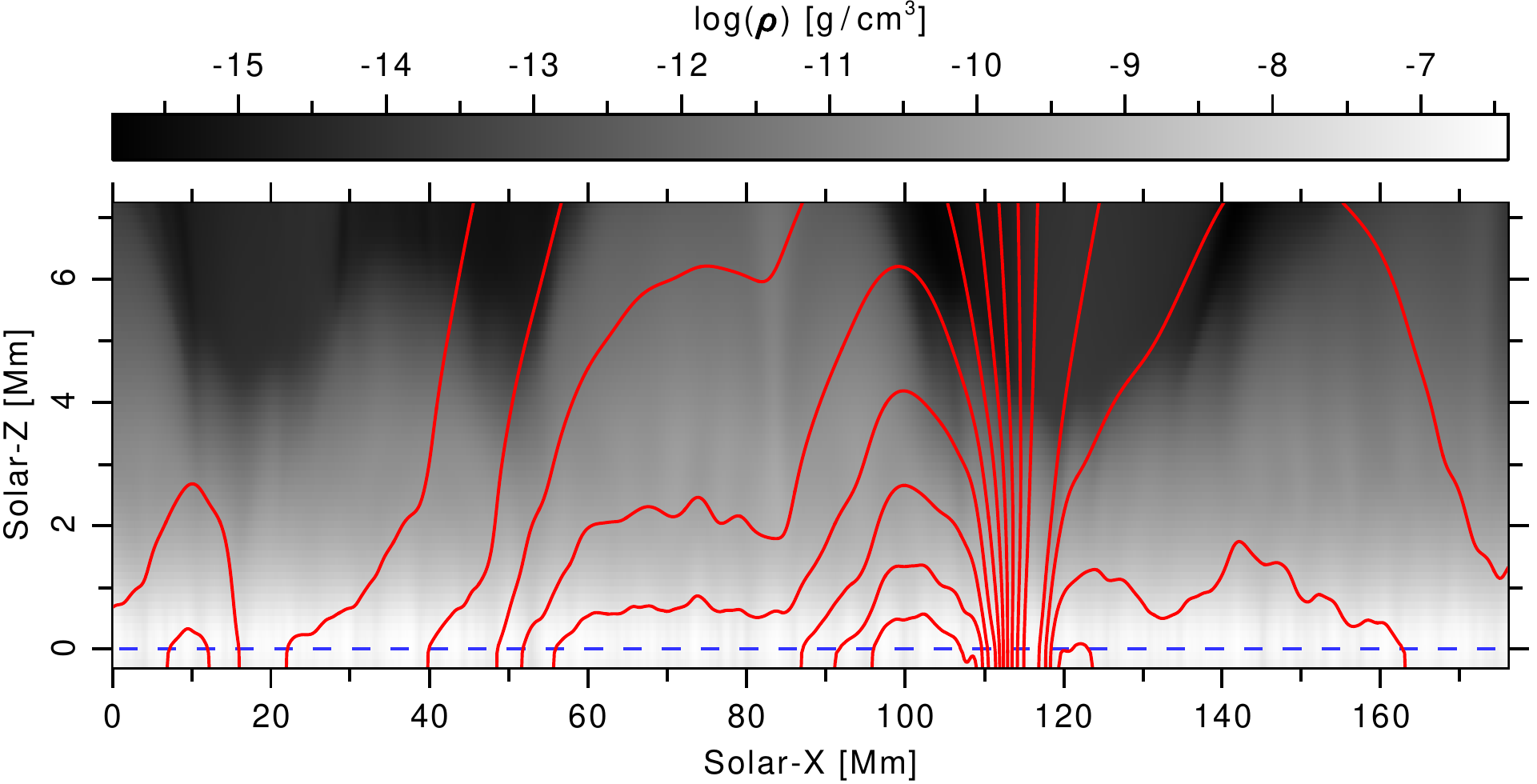}{field-lines_xz-cut}{In-plane magnetic field lines (red) in a vertical cut through a strong flux concentration in the lower part of the model. The gray scale indicates the logarithmic density. The blue dashed line is the actual physical boundary of the simulation domain at \eqi{z = 0\unit{Mm}}.}

In order to mimic the increase in atmospheric and magnetic pressure in the solar interior below the photosphere, we use an inverted potential-field extrapolation that concentrates any magnetic flux bundles with depth, which leads to a reduction of the diameter and an increase of the amplitude of those flux bundles.
Typically, the pressure scale height below the photosphere is about \eqi{300\unit{km}}.
As we use three ghost layers below the lower boundary with a grid distance of about \eqi{\Delta z = 100\unit{km}}, we would have to double the contrasts from the surface magnetograms observed at \eqi{z(0) = 0\unit{Mm}}.
This is not possible without introducing artifacts like wiggles and checker-board patterns in the ghost cells.
Therefore, we reduce the contrast increase in the lower ghost cells by a constant factor, here to one fifth, as if the pressure scale height would be about \eqi{1.5\unit{Mm}}; see also \cite{Bourdin+al:2018_helicity} for a more mathematical description.
Of course, this reduction introduces a slight error in the horizontal field components below the surface.
But this error seems acceptable, because significant magnetic structures in the photosphere typically have a horizontal diameter of well above \eqi{1\unit{Mm}} and the field therein is anyway mainly vertical for strong flux concentrations that finally may reach our model corona; see \fig{field-lines_xz-cut}.

In principle, one could say that this description of the lower magnetic field boundary is similar to a strictly vertical field boundary condition.
We like to note this is not the case, because we still allow for any amount of horizontal field at the boundary and hence very small near-surface closed field lines may form; see red lines near \eqi{x = 10} and \eqi{120\unit{Mm}} in \fig{field-lines_xz-cut}, where at the latter position a field line even gets horizontal below the actual boundary.
Such cases are rare, which supports our argument that the error is small when enlarging the pressure scale height as described above, but this shows we do not enforce a vertical field in the photosphere and below.

Different and less complex boundary conditions for the magnetic field may of course exist.
For example, we could enforce a vertical field at the lower boundary.
But when we do that, many field lines in the physical domain that are not strictly vertical, would have a stronger kink near the boundary as with a more flexible boundary condition.
An enforced vertical field at the bottom will generate larger derivatives and hence artificial currents near the photosphere for a multi-polar AR case.
In comparison, the vertical-field boundary condition has about twice the current density, twice the maximum magnetic Reynolds number, and about \eqi{40\unit{\%}} larger Lorenz forces than we find with the more flexible scheme we use here at the bottom of the domain.
This would then require us to double our magnetic diffusivity \eqi{\eta} in order to keep the magnetic Reynolds number under control.
As a consequence, the larger diffusivity leads to more slippage of the field lines when we advect them with our observational driving --- which means the driving looses much of its effect.
Therefore, currents at the bottom boundary are unwanted and we like to keep the magnetic diffusivity minimal to get the model more realistic.

Previously, the {\em{Pencil Code}} had also a regular potential-field extrapolation that relaxes the fields to a force-free state outside the physical domain.
This would mean to smear out constrasts in the magnetic field below the photosphere.
For the top boundary this is acceptable, because the field is already quite force free, there.
But below the photosphere, the pressure increases and magnetic structures are not force free.
If we would apply the original potential-field extrapolation, we again generate strong currents because the field will relax to a force-free state.
This is a permanent process, because we constantly keep on driving the field away from that state.
Therefore, our method with increasing magnetic contrasts in the ghost cells below the bottom boundary is closer to reality than using the regular potential-field extrapolation.

At the top boundary, we may use the regular potential-field extrapolation with contrasts that get smeared out exponentially with height, which is fairly realistic in the outer corona.%
\footnote{In the \texttt{start.in} configuration file we set \texttt{bcz} to \texttt{\apos{pfe}} for the first magnetic field component. At the same component's position we also set \texttt{fbcz\_bot} only for the bottom boundary to \texttt{0.2} in order to limit the extrapolation below the photosphere to one fifth of the pressure scale height.}

\subsection{Atmospheric boundary condition}

At the lower boundary and below (in the ghost layers) we prescribe the density according to a smoothly combined profile of the stellar interior and the solar atmosphere, as given in \cite{Bourdin:2014_switch-on}.
We use a globally constant diffusivity of \eqi{D_\rho=8 \cdot 10^6\unit{m^2/s}} in the mass density \eqi{\rho}, which is needed for the numerical stability in all independent MHD variables.
This results in a very small mass transport into the simulation box through the lower boundary because the diffusion acts along the gradient that points upwards.
Implicitly, the thermal pressure increases in the upper layers.
This pressure increase gets exactly compensated by a hydrodynamic flow equilibrium that requires a downwards bulk plasma motion.
Consequently, the photospheric density at the physical boundary layer may vary slightly from the preset value.
The closed \eqi{z}-boundary condition sets the vertical bulk velocity \eqi{u_z = 0} at the lower physical boundary \eqi{z(0)}.
Hence, a mass inflow would then accumulate.
To compensate for continuous density changes due to the diffusion, we have to constantly overwrite the density to its initial profile at and below the photosphere.
The layers above the photosphere undergo the regular hydrodynamic settlement of the atmospheric stratification due to the gas pressure, mass flows, and gravity.

We set the temperature boundary condition so that a consistent hydrostatic equilibrium is reached at the lower boundary.
To achieve this, we recursively set the temperature \eqi{\left. T \right|_j} in the three lower ghost layers \eqi{j \in \{-1,-2,-3\}} at the grid positions \eqi{z(j)} as:%
\footnote{In the \texttt{start.in} configuration file we set \texttt{bcz} to \texttt{\apos{fg}} for the density and to \texttt{\apos{hse}} for the temperature.}
\eql{\left. T \right|_j = \left. T \right|_{j+1} ~ {\left. \rho \right|_{j+1} \over \left. \rho \right|_j} ~ \exp \left [- \left. \Delta z \right|_j ~ {c_v \over c_p} ~ {\left. g \right|_{j+1} \over \left. T \right|_{j+1}} \right ] , \label{E:T_lower}}
where we use the identity \eqi{c_v = \gamma/(\gamma-1)} for an ideal gas and \eqi{\left. \Delta z \right|_j = z(j+1) - z(j) > 0} for the bottom boundary.
\eqi{\left. \rho \right|_j} is the density and \eqi{\left. g \right|_j} is the gravity constant at the grid positions \eqi{z(j)}.



An alternative, without prescribing the density to its initial value, is to maintain a zero first derivative of the temperature by a symmetric boundary condition and then adapt the density according to a hydrostatic equilibrium.
This has the advantages that, first, there is no heat flow into or out of the simulation domain, and second, that the density may deviate from its initial value and maintain a hydrostatic equilibrium at the boundary; see \sect{closed.atmosphere} for details.

\section{Chromospheric nudging}

Radiative losses act as an energy sink in the solar atmosphere \citep{Cook+al:1989}.
The Spitzer heat conduction transports energy along the temperature gradient \citep{Spitzer+Härm:1953} and therefore provides an energy source in the chromosphere.
Both together may destabilize the chromosphere because once the heat conduction provides less energy, the radiative losses cool the plasma that hence becomes denser.
Denser plasma is a stronger source of radiation that cools the plasma further.
This may form a run-away effect, because there is no radiative source term in our model that would heat optically thicker plasma through absorption.

Another run-away effect exists for an excess heat input that leads to an adiabatic expansion of some plasma which then looses its ability to radiate an excess of internal energy, because the radiative losses decrease for lower densities.
Therefore, we need to stabilize our model chromosphere with a so-called Newtonian cooling method, which resembles the stabilizing effects of chromospheric radiative transfer at low computational costs.

\subsection{Newtonian cooling}\label{S:newton.cool}
Temperature fluctuations in the lower solar atmosphere get smoothened out through the radiative transfer that efficiently heats denser and optical thicker plasma, while it also cools less dense plasma through reduced absorption and the continuous emission of photons \citep{Spiegel:1957}.
Both stabilizing effects of a proper radiative transfer treatment in the chromosphere can be achieved through a novel combined Newtonian cooling and chromospheric nudging approach.

First, we gently push back the chromospheric temperature to its initial stratification \eqi{T_0(z)} through an exponential decay.
Second, we implement a cooling term that slowly pushes the temperature to a target value \eqi{T'(\rho({\vec{r}},t))} at the position \eqi{{\vec{r}}=(x,y,z)} that we take from the initial temperature stratification \eqi{T_0(z')} at the height \eqi{z'} where the actual density \eqi{\rho({\vec{r}},t)} is equal to the initial density stratification \eqi{\rho_0(z') = \rho({\vec{r}},t)}.
We finally obtain the target temperature \eqi{T^\ast} for the nudging as:
\eql{T^\ast({\vec{r}},t) = \sqrt{T_0(z) T'(\rho({\vec{r}},t))}}

Of course, we like to apply the Newtonian cooling only in the chromosphere, where our model needs it, and not beyond.
This we achieve by a cutoff mechanism with four components: 1) we use a density-dependent cutoff that sets in when the density becomes eight orders of natural-logarithmic magnitude smaller than the photospheric value, 2) we smooth out this sharp cutoff boundary with a sine function over the last two orders in this natural-logarithmic magnitude, 3) we enforce another cutoff that sets in above a height of 3\unit{Mm}, and 4) we apply another smooth sine transition over the upper 0.3\unit{Mm} on this height-dependant cutoff.

We now add the nudging term to the energy balance (shortened by `...'):%
\footnote{The related parameters in the \texttt{solar\_corona} module are the logical flag \texttt{lnc\_density\_depend} to activate the density-dependent Newtonian cooling, \texttt{nc\_tau} as the inverse time scale necessary for the nudging by an exponential decay, \texttt{nc\_lnrho\_num\_magn} as the number of natural-logarithmic magnitudes in density, \texttt{nc\_lnrho\_trans\_width} defining the smooth transition of the density-dependent cutoff, \texttt{nc\_z\_max} as the maximum height to apply the Newtonian cooling, and \texttt{nc\_z\_trans\_width} to define the smooth transition for the height-dependent cutoff.}
\eql{\partial T / \partial t = ... + T \exp \left [ -\tau_t \left ( 1 - {T^\ast \over T} \right ) ~ c_{\rho} ~ c_{z} \right ]}
with the inverse decay half-time \eqi{\tau_t = 0.5\unit{s^{-1}}}, \eqi{c_{\rho}} as the smooth density-dependent cutoff function, and \eqi{c_{z}} as the smooth height-dependent cutoff.

\subsection{Compressible atmospheric column}
\graphfull{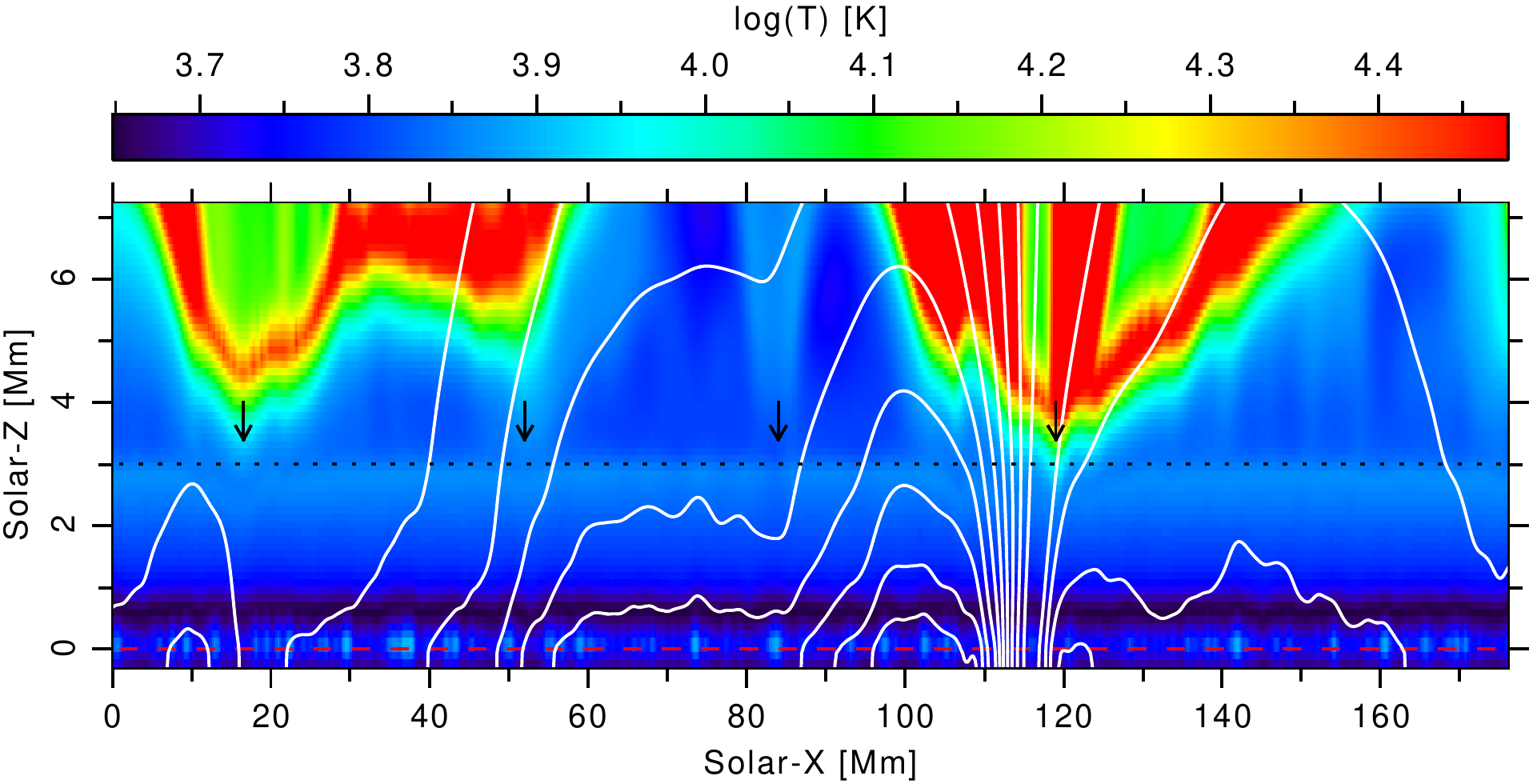}{Temp_xz-cut}{Logarithmic temperature (color code) in the lower atmosphere for a vertical cut through a strong flux concentration together with the in-plane magnetic field lines (white) at the same location as in \fig{field-lines_xz-cut}.
The red-dashed line indicates the location of the photosphere.}

The effect of the Newtonian cooling becomes visible in \fig{Temp_xz-cut}, where we see a relatively constant chromospheric temperature is maintained during the simulation run and remains similar to the initial condition below 3\unit{Mm} (black dotted).
We indicate locations with down arrows, where we find that the atmospheric column gets compressed and hence the temperature follows a similarly compressed stratification.
With time, this atmospheric column compression will relax towards the initial state due to our combined Newtonian nudging method; see \sect{newton.cool}.
This allows our model chromosphere to adapt to, e.g., downflows from the corona and eventually relax back to the initial stratification after these downflows end.

At some locations we find the atmospheric column rises quickly to coronal temperatures above 3\unit{Mm}; see leftmost and rightmost down arrows in \fig{Temp_xz-cut}.
In other regions, the lower-coronal plasma remains relatively cool, typically above regions with more horizontal field; see at Solar-X from 60 to 95\unit{Mm}.

In the photosphere, we see that temperatures may be higher and lower, where the density is lower and higher, respectively.
This anti-correlating behavior of temperature and density could be due to some compressional heating from the granular horizontal advecting motions; see the lower boundaries in \fig{field-lines_xz-cut} and \fig{Temp_xz-cut} at \eqi{z = 0\unit{Mm}} (dashed lines).
Also these local changes in temperature will eventually relax to the initial photospheric temperature due to the limited lifetime of granules and the Newtonian cooling.

We find also that the height of the photospheric temperature minimum varies between about 250 to 350\unit{km} due to our compressible atmospheric column at the lower boundary of our simulation domain; see wavy black layer in the lower part of \fig{Temp_xz-cut}.

\section{Upper coronal boundaries}

On the upper end of the simulation domain, we prefer not to allow heat or plasma inflows or outflows.
The reason is that any inflow would be of undefined velocity, temperature and density.
We would anyway not expect heat outflows, because the temperature gradient usually leads only to inflows of heat.
At the same time, such inflows are unwanted, because we like to learn about the intrinsic heating of the corona, independent of any boundary condition.

If we allow for plasma outflows, this may lead to an unrealistic mass loss, because in reality this outflowing plasma may well fall back to the Sun, later, due to gravity.
Therefore, we prefer to make the simulation domain large enough to capture all relevant plasma flows and to simply close the upper simulation boundary for any plasma and heat flows.

For the magnetic field, though, we like to allow for ``open'' field lines instead of formulating a boundary condition that either enforces vertical or horizontal fields at the top boundary; as we describe in \sect{potential.field.extrapolation} below.

\subsection{Closed atmospheric boundary} \label{S:closed.atmosphere}

Above the upper boundary at 156\unit{Mm} the coronal temperature in standard 1D atmospheric stratifications would still rise \citep{Bourdin:2014_switch-on}.
If we simply prescribe the temperature in the upper ghost layers from such a stratification, we would impose an unwanted thermal energy inflow downwards into the physical box due to heat conduction along the temperature gradient.
Therefore, and in contrast to the lower atmospheric boundary in the photosphere, we need to employ a different boundary condition at the upper coronal boundary (in grid cell \eqi{nz}) that assures there is no heat inflow.
We achieve this by forcing the temperature gradient to be zero at the upper physical boundary (\eqi{\left. z \right|_{nz} = 156\unit{Mm}}) with a symmetric boundary condition for the three upper ghost cells \eqi{j \in \{1,2,3\}} at the grid positions \eqi{\left. z \right|_{nz+j}} as:
\eql{\left. T \right|_{nz + j} = \left. T \right|_{nz - j} \label{E:T_upper}}
Since the temperature near the boundary is almost uniform, also inclined field lines see a symmetric temperature stratification.

We then need to set the density \eqi{\left. \rho \right|_{nz + j}} again consistent with a hydrostatic equilibrium:
\eql{\left. \partial p \over \partial z \right|_{nz + j} = \left. \rho \right|_{nz + j} \left. g \right|_{nz + j}}
This leads us to an upper boundary condition for the density that would require a constant temperature, constant gravity, and symmetric grid distances at the top:%
\footnote{In the \texttt{run.in} configuration file we set \texttt{bcz} to \texttt{\apos{hs}} for the density and to \texttt{\apos{s}} for the temperature.}
\eql{\Rightarrow \left. \rho \right|_{nz+j} = \left. \rho \right|_{nz-j} \exp \left[- \frac{1}{\gamma} \left( \left. z \right|_{nz+j} - \left. z \right|_{nz-j} \right) {{\left. g \right|_{nz}} \over {\left. T \right|_{nz}}} \right] \label{E:rho_upper}}

For a non-constant temperature, non-constant gravity, and arbitrary grid distances at the boundary, the density in the ghost layers can be formulated in a recursive way as:
\eql{\Rightarrow \left. \rho \right|_{nz+j} = \left. \rho \right|_{nz+j-1} \exp \left[- \frac{1}{2 \gamma} \left( \left. \Delta' z \right|_{nz+j-1} {{\left. g \right|_{nz+j-1}} \over {\left. T \right|_{nz+j-1}}} + \left. \Delta' z \right|_{nz+j} {{\left. g \right|_{nz+j}} \over {\left. T \right|_{nz+j}}} \right) \right] , \label{E:rho_upper_general}}
where \eqi{\left. \Delta' z \right|_j} denotes the vertical extent of the grid cell at \eqi{z(j)}.
If the grid spacing is close to being equidistant near the boundary, we may simplify \eqn{rho_upper_general} to:%
\footnote{For non-constant gravity in the file \texttt{run.in} we set \texttt{bcz} to \texttt{\apos{hse}} for the density and to \texttt{\apos{s}} for the temperature.}
\eql{\Rightarrow \left. \rho \right|_{nz+j} = \left. \rho \right|_{nz+j-1} \exp \left[- \frac{1}{\gamma} \left( \left. z \right|_{nz+j} - \left. z \right|_{nz+j-1} \right) {{\left. g \right|_{nz+j} + \left. g \right|_{nz+j-1}} \over {\left. T \right|_{nz+j} + \left. T \right|_{nz+j-1}}} \right] . \label{E:rho_upper_general_equidist}}


\subsection{Potential-field extrapolation} \label{S:potential.field.extrapolation}

When the simulation domain is large enough, the magnetic field near the top boundary should have reached a nearly potential state.
We may then use a purely potential field in the three ghost layers above the domain as boundary condition for the magnetic field.
Because the field vectors above and below the upper boundary are not perfectly rotation free, we still obtain some currents at the physical boundary, which represent the relaxation from a nearly to a fully potential state.
If these currents are strong, they might artificially heat the corona.
To counter this effect, we employ an additional magnetic diffusivity below the boundary to smoothen the transition to the potential state and reduce these artificial currents, as we describe in \sect{mag.swamp} below.

For the potential-field extrapolation into the ghost layers, we Fourier-transform the magnetic vector potential \eqi{\vec{A}}, extrapolate it by smoothing out contrasts, and then transform it back:
\eql{\left. \widehat{\vec{A}}(k_x,k_y,t) \right|_{nz} = \int \left. \vec{A}(x,y,t) \right|_{nz} ~ e^{{\rm{i}}\vec{k} \cdot \vec{r}} ~ {\rm{d}}^2\vec{r} ~ ,}
where the vector \eqi{\vec{r} = (x,y)} lies in the horizontal plane and \eqi{\vec{k} = (k_x,k_y)} denotes the horizontal wave vector.
We extrapolate from the location of the physical boundary \eqi{z(nz)} to the coordinates \eqi{z(nz+j)} of the three ghost layers \eqi{j \in \{1,2,3\}} with:%
\footnote{In the \texttt{run.in} configuration file we set \texttt{bcz} to \texttt{\apos{pfe}} for the first magnetic field component, which also sets the other two components accordingly.}
\eql{\left. \widehat{\vec{A}}(k_x,k_y,t) \right|_{nz+j} = \left. \widehat{\vec{A}}(k_x,k_y,t) \right|_{nz} ~ \exp \left[ - |\vec{k}| \left. \Delta'' z \right|_{nz+j} \right] \label{E:potential_field_extrapolation}}
We denote here the distance between the physical boundary and the extrapolated layer as \eqi{\left. \Delta'' z \right|_{nz+j} = z(nz+j) - z(nz) > 0}.
The normalized Fourier back transform of \eqi{\widehat{\vec{A}}} finally gives us the vector potential \eqi{\vec{A}} in the upper three ghost layers.
Because \eqi{\left. \Delta'' z \right|_{nz+j}} is positive on the upper boundary, we smear out any contrasts in \eqi{\vec{A}} with increasing height.

\subsection{Diffusive swamp region}

The potential-field extrapolation relaxes the magnetic field very quickly into a force-free state and hence strong currents may emerge at the upper physical boundary.
Also a strong heat transport towards the upper simulation domain may become problematic, because the upper boundary is closed for heat flows and we may accumulate very high temperatures that become numerically demanding in the low-density plasma of the outer corona, where this plasma can not radiate an excess of internal energy.
To circumvent both problems, we use an additional diffusivity for the magnetic field and for the heat conduction below the physical boundary by implementing a so-called swamp region.

The swamp diffusivities act only in the upper part of the simulation domain to reduce the artificial effects on the upper boundary.
We achieve a smooth transition by a height-dependent weighting function \eqi{w(z)} that smoothly goes from 0 below to 1 within the swamp region.
For this transition we use a cubic step function that starts at height \eqi{z_{s}} and ends at \eqi{z_{t}}.
The swamp region is fully active above \eqi{z_{t}} so that \eqi{w(z \leq z_{s}) = 0} and \eqi{w(z \geq z_{t}) = 1} with zero first derivatives at the edges of the transition function \eqi{w(z)}.%
\footnote{In the \texttt{run.in} configuration file the parameters \texttt{swamp\_fade\_start} and \texttt{swamp\_fade\_end} define the height of the smooth transition to the swamp region.}


\subsubsection{Magnetic diffusivity} \label{S:mag.swamp}
\noindent
To diffuse out currents near the upper end of the simulation domain, we implement an additional isotropic magnetic diffusivity within the swamp region.
This swamp diffusivity is of course omitted in the energy balance in order not to heat the corona artificially.
We multiply the weighting function \eqi{w(z)} to the constant swamp diffusivity \eqi{\eta_{s}} and obtain the height-dependent magnetic swamp diffusivity \eqi{\eta_{s}(z) = \eta_{s} ~ w(z)}.
Now we add the magnetic swamp diffusion term to the induction equation (shortened by `...'):%
\footnote{In the \texttt{run.in} configuration file one may activate the magnetic swamp diffusivity by setting \texttt{swamp\_eta} to a value larger than zero and similar to the parameter \texttt{eta}.}
\eql{{\partial \vec{A} \over \partial t} = ... + \eta_{s}(z) ~ \Delta \vec{A} + \vec{e}_z ~ {\partial\eta_{s}(z) \over \partial z} ~ \vec{\nabla} {\bf \cdot} \vec{A}}

In the case of \cite{Bourdin+al:2013_overview} the simulation domain was large enough, so that currents at the top boundary were not problematic and hence no magnetic swamp diffusivity was used there.

\subsubsection{Heat conduction}
\noindent
Similar to the magnetic swamp diffusivity (see above), we implement also a diffusivity that acts on the temperature as a constant, uniform, and isotropic heat conduction.
For that, we add another term to the energy balance (shortened by `...') and use \eqi{\chi_{s}} as the swamp heat diffusivity constant:%
\footnote{In the \texttt{run.in} configuration file we set the parameter \texttt{swamp\_chi} larger than zero to activate the heat swamp diffusion.}
\eql{{\partial T \over \partial t} = ... + \chi_{s} ~ w(z) ~ \Delta T \label{E:swamp_chi}}
Again, the height-dependent function \eqi{w(z)} provides a smooth transition between the physical regime and the swamp region.

We like note that \eqi{\chi_{s}} acts either on the temperature \eqi{T}, the natural-logarithmic temperature \eqi{\ln T}, or the entropy \eqi{\epsilon}, depending on which quantity is active in the simulation setup.
Therefore, the physical unit of \eqi{\chi_{s}} is not identical across these cases and may be different from the units of \eqi{\chi} used for the regular global isotropic heat conduction \eqi{\chi ~ \Delta T}.

\subsubsection{Mass diffusion}
\noindent
Similar as for the heat conduction, we implement another swamp region that acts on the density with a diffusive term added to the continuity equation (shortened by `...'):%
\footnote{We activate the density swamp region via the \texttt{swamp\_diffrho} parameter in the \texttt{run.in} configuration file.}
\eql{{D\rho \over Dt} = ... + \chi_{\rho} ~ w(z) ~ \Delta \rho}
Like for \eqn{swamp_chi} we use the same diffusivity parameter \eqi{\chi_{\rho}} with different physical units for either the density or the natural-logarithmic density, depending on what quantity is used for the simulation.

\section{Massive-parallel methods}

For high-performance computing applications it is crucial to reduce all computations that are serialized or restricted to few processors.
The parallelization of such computations therefore helps to scale an application efficiently to substantially more processors.
In the following sections we describe some massive-parallel methods introduced to the {\em{Pencil Code}} in the years from 2009 to 2013.

One bottleneck in massive parallelization for simulation runs like described in \cite{Bourdin+al:2013_overview} is the potential-field boundary condition that uses a Fast Fourier Transform (FFT), as well as the massive-parallel file input and output described in \sect{IO}.

\subsection{Fast Fourier Transform} \label{S:FFT}

Some boundary conditions, like a potential-field extrapolation, need to compute the FFT along both horizontal directions.
This requires to collect once all data along \eqi{x} and then once along \eqi{y}.
Less parallelized FFT routines collect all data on one processor, perform the FFT, and then distribute the transformed data back to all processors.
During the computationally expensive communication and computation, a large number of processors are idle and their potential resources remain unused.

A more efficient method is to collect all data along the one direction which the FFT actually needs and to split the domain along the other direction; see the remapping scheme presented in \fig{FFT-remapping}.
As a result, all processors in one \eqi{(x,y)}-layer may contribute in parallel to the computation of the FFT along one direction.
Furthermore, the remapping of the data can be done in a parallel way, so that all communication finishes after only two communication cycles: one send and one receive operation.
The trick is to let one half of the processors send first and then receive, while the other half first receive and then send their local data portion.
This is actually faster than to collect all data on one processor, because this can only be done in a sequential way and requires significantly more communication cycles for a large number of processors.
The same holds true for the remapping back to the original subdomains.

\graphwidth{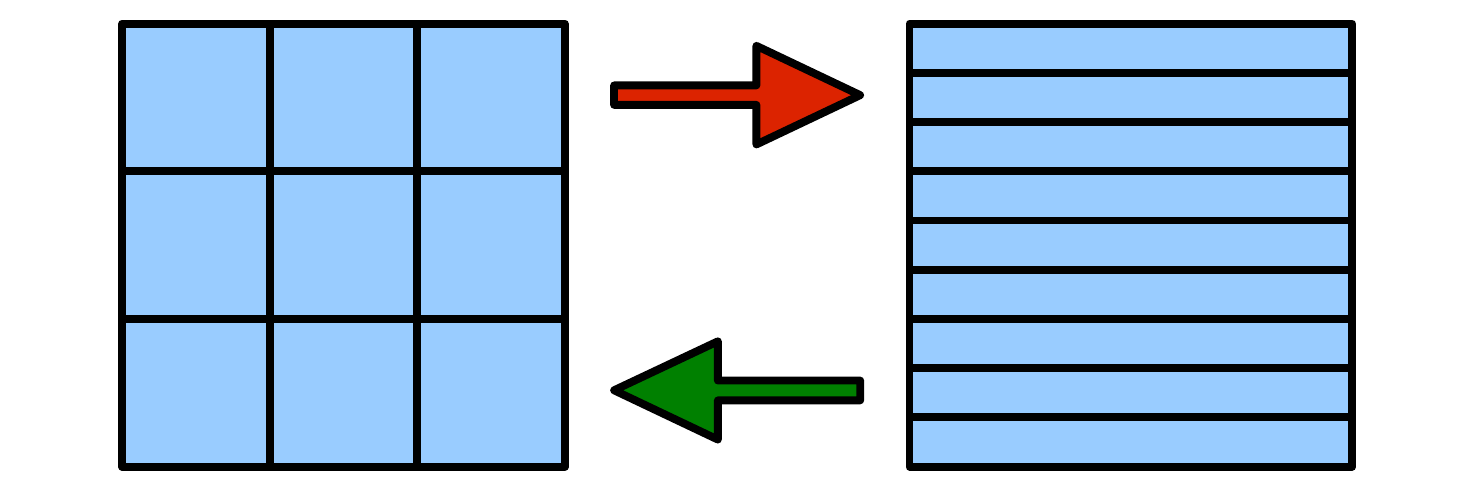}{FFT-remapping}{Data remapping strategy for the parallel FFTs in the module \texttt{`fourier\_fftpack'}.}

The required transpose operation of the remapped data during a two-dimensional FFT needs more communication cycles, namely as many as the number of subdomains along the \eqi{x}-direction.
Still, this number is typically less than the number of processors in one \eqi{(x,y)}-layer and the transpose can be parallelized between the processors in the \eqi{y}-direction.

Due to the participation of many processors in the FFT, the whole computation, including the multiple data remappings, is faster than the original scheme of the FFT for only one processor.
When the domain is not split into subdomains along \eqi{x}, both schemes are of course similar, but this is usually not the case for large simulation runs with many processors.

\subsection{Input/Output} \label{S:IO}

One challenge for large-scale simulation runs is the input and output (IO) of data snapshots, as well as their storage requirements.
With clever IO strategies one can not only perform IO operations faster, but also save significant amounts of storage space.

Traditionally, the {\em{Pencil Code}} stores data snapshots in a `distributed' manner, where each processor writes one file with its local subdomain portion of the data, like implemented in the \texttt{`io\_dist'} module.
This strategy has the advantage that the writing can be done fully in parallel, but at the cost of writing all inner ghost layers between any neighboring processors, which contains overlapping and identical data.
The smaller the subdomains get and the more processors we like to use in order to boost the computation speed, the larger is the amount of unneeded data that this distributed method requires to store.
Also a fully distributed IO method will cause problems on any file system, because these are not made for thousands or more simultaneous IO requests.

\graphfull{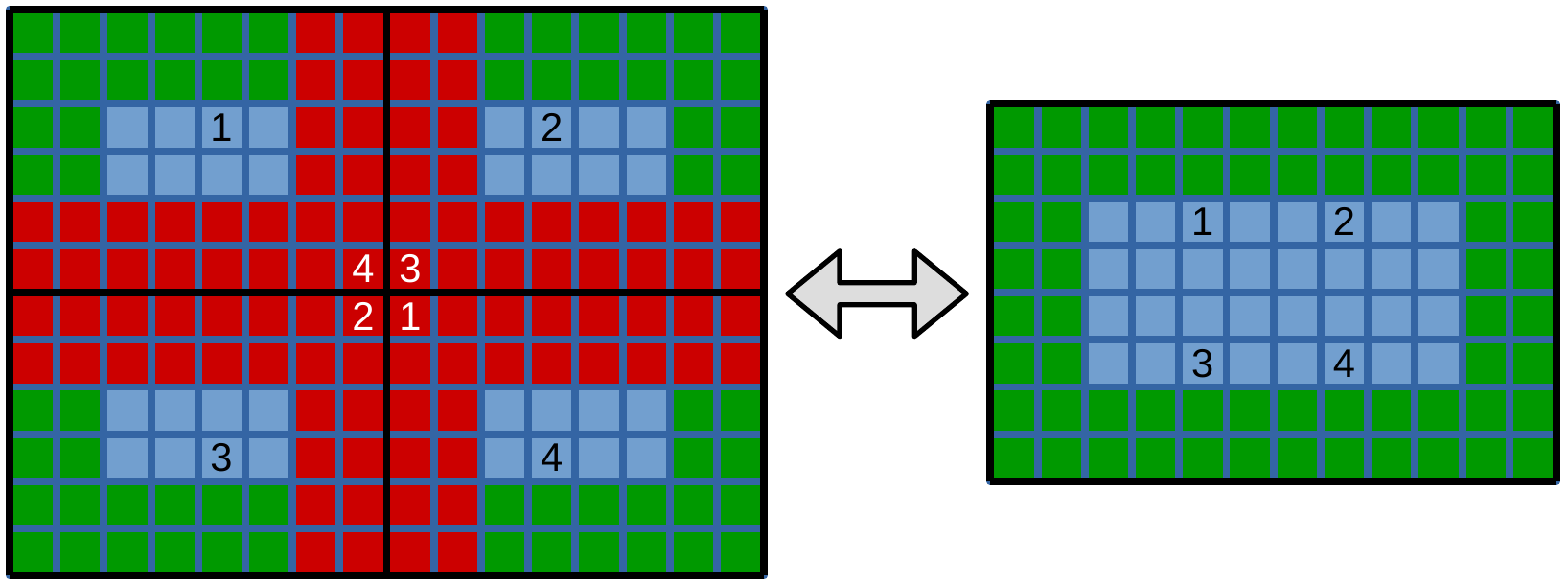}{ghosts}{Reduction of required data space for monolithic files in an extreme case, where the needed outer ghosts are colored in green and the physical domain is blue. The redundant inner ghost cells (see numbers) being saved with the distributed \texttt{`io\_dist'} IO module in the {\em{Pencil Code}} are highlighted in red. Black lines depict the data boundaries for each file. There number of ghost layers is 2 for this example.}

\graphwidth{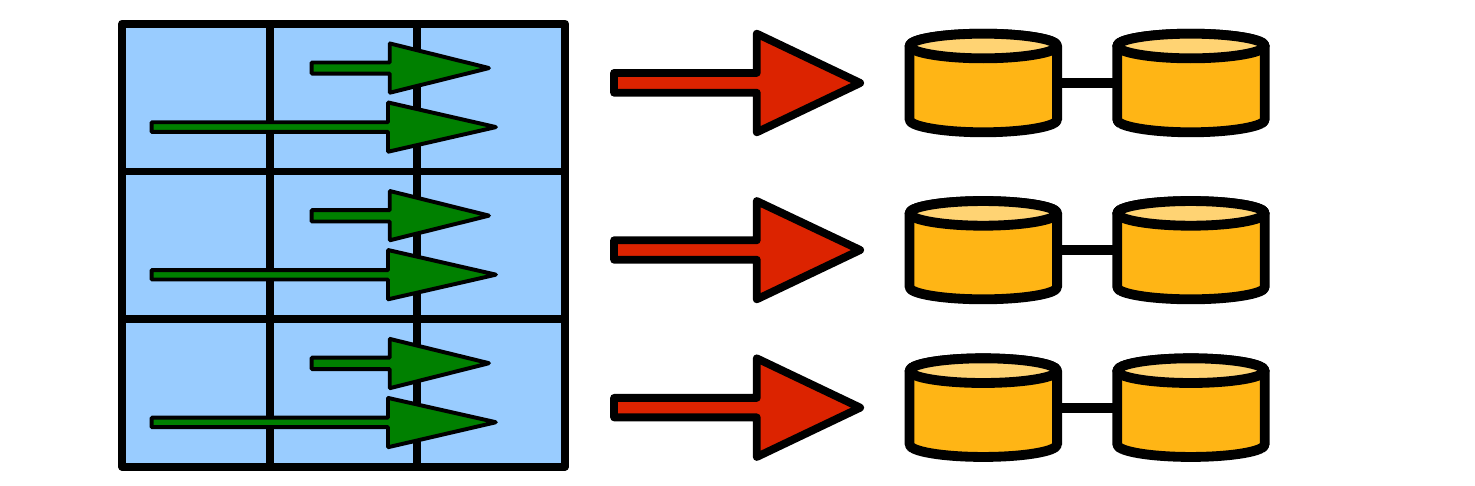}{IO-strategy}{Collective IO strategy of the module \texttt{`collect\_xy'}, where data is collected in subdomain layers along \eqi{(x,y)} and is written to multiple files by the collecting (here rightmost) processors in each \eqi{(x,y)}-plane. The vertical separation of the layers depicts the \eqi{z}-direction in the simulation domain.}

In a first step, one might try to collect all data on one processor and write it out into one monolithic snapshot file.
This strategy we call `collective' and it is implemented in the \texttt{`io\_collect'} module.
It turns out this method is not optimal regarding the IO speed, because one processor alone can only access a data snapshot in a sequential way.
Nonetheless, we may omit to store all inner ghost cells; see the scheme displayed in \fig{ghosts}.

The next step of improving the IO lies in combining the distributed with the collective strategy.
This we implement as the \texttt{`io\_collect\_xy'} module, where all data is firstly collected in along \eqi{(x,y)}-layers by the leading processor in this layer and secondly written by all \eqi{(x,y)}-leading processors in parallel; see \fig{IO-strategy}.
The reading access works in the same way: the \eqi{(x,y)}-leading processors read in parallel and then distribute the data within their layer.
For this improvement one has to take the disadvantage of storing the inner ghost cells between all \eqi{(x,y)}-layers, which is still significantly less than storing all inner ghost cells.
Hence, this advanced method has substantial potential to accelerate the IO by some parallelization.

Most modern IO methods usually use an intermediate software layers to optimize the number of parallel IO requests and to write out monolithic files that do not need to store any inner ghost cells at all.
Such file formats therefore have the potential to save a substantial amount of storage space.
In the {\em{Pencil Code}} we now provide two IO modules: \texttt{`io\_mpi2'} that relies on the {\em{MPI-2}} standard and \texttt{`io\_hdf5'} that uses the parallel {\em{HDF5}} software library for IO operations.
Both use monolithic file formats and hence are optimal regarding the data storage requirements.
In the same time, their IO routines are also optimized for scalability and speed; see comparison in \tab{timings}.

\subsection{{\em{HDF5}} file format} \label{S:HDF5}

\begin{table}
\centering
\begin{minipage}[t]{12cm}
\caption{Properties of the different IO strategies in {\em{Pencil Code}} for writing a full data snapshot.}
\label{T:timings}
\tablefont
\begin{tabular}{lrcl}
\toprule
module		& time				& storage			& notes \\ 
\colrule
distributed	& 2\unit{s}			& \bfseries\red{+31\unit{\%}}	& \bfseries\red{not scalable above 256 CPUs} \\
collect		& \bfseries\red{70\unit{s}}	& min.				& one IO node collects globally \\
collect\_xy	& 10\unit{s}			& \bfseries\red{+16\unit{\%}}	& one IO node per \eqi{(x,y)}-layer \\
mpi2		& 8\unit{s}			& min.				& \bfseries\red{binary IO hidden in MPI2} \\
hdf5		& 9\unit{s}			& min.				& portable, extendable structure \\
\botrule
\end{tabular}
\end{minipage}
\end{table}

The {\em{Pencil Code}} is now capable of storing monolithic data snapshots with minimal storage requirements in the {\em{HDF5}} format.
Snapshot files (that were previously in a raw binary format) are now stored in the self-explanatory {\em{HDF5}} format with an extendable data structure.
All {\em{HDF5}} files carry the filename suffix \texttt{`.h5'} instead of the binary \texttt{`.dat'} file extension.

Once a user switches to the {\em{HDF5}} file format, the build process (\texttt{pc\_build} or \texttt{make}) tries to automatically find the {\em{HDF5}} libaray and uses the location where the Fortran compiler wrapper is present.%
\footnote{We switch to the {\em{HDF5}} format by setting \texttt{IO = io\_hdf5} and \texttt{HDF5\_IO = hdf5\_io\_parallel} in \texttt{src/Makefile.local}.
On a standard \texttt{ubuntu 18.04 LTS} system, one needs to install the package \texttt{`libhdf5-openmpi-dev'}.
The packages \texttt{`hdf5-tools'} and \texttt{`hdfview'} contain optional tools for inspecting and modifying {\em{HDF5}} files.}
For that, the \texttt{`\$PATH'} environment variable must point to one of those compiler wrappers (\texttt{h5pfc} or \texttt{h5fc}).
Otherwise, no configuration by the user is required.
In a computing center, this usually requires to load an environment module with the command \texttt{`module load {\dots}hdf5{\dots}'}, where all available modules can be listed with \texttt{`module avail'}.

\graphfull{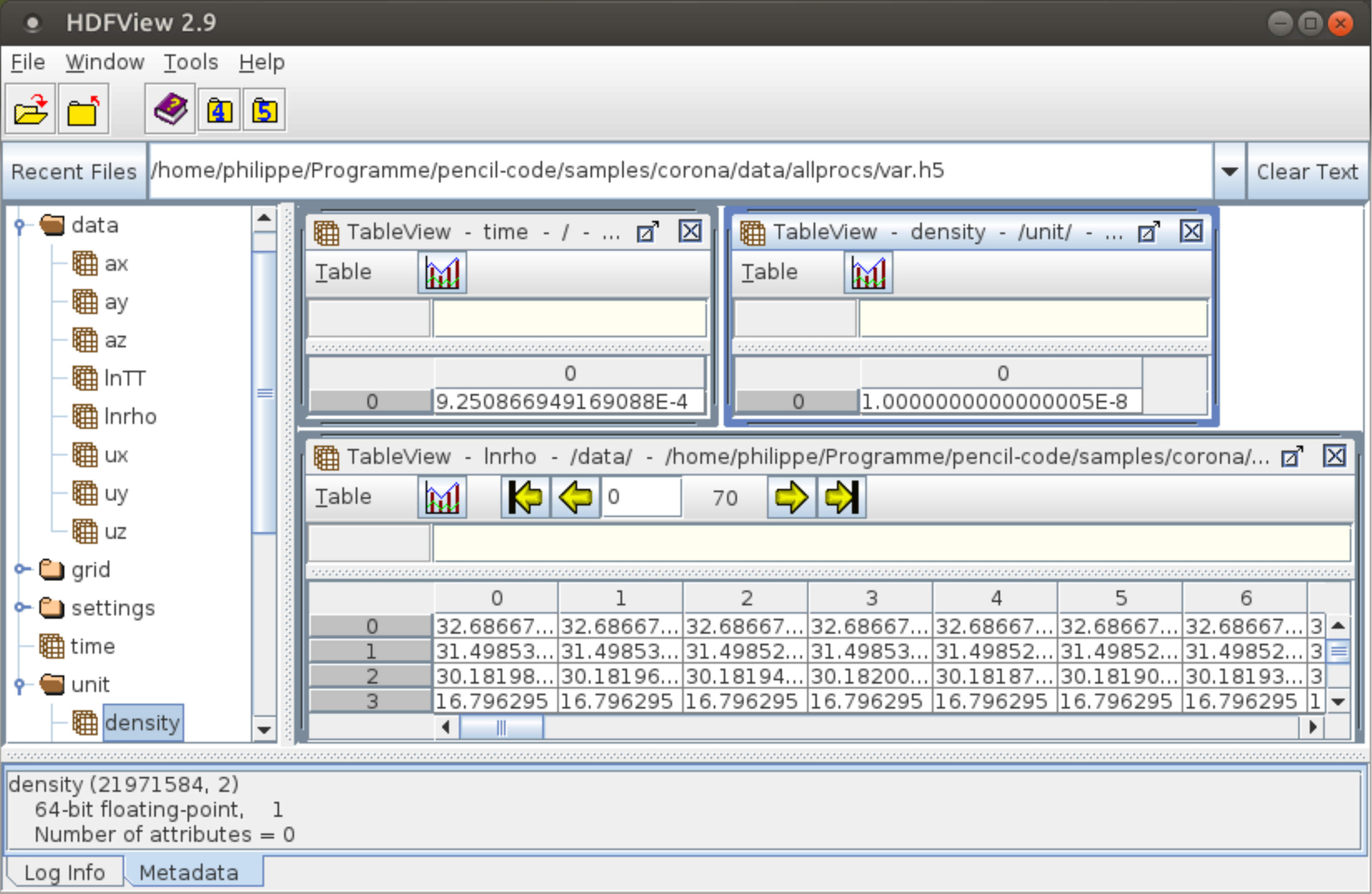}{varfile}{Example content of a \texttt{`var.h5'} file, here from the \texttt{corona} sample. The conversion factor from code units to physical units is \eqi{10^{-8}} for the density. This sample stores the logarithmic density \texttt{lnrho} as three-dimensional data array, together with the logarithmic temperature \texttt{lnTT}, and all components of the vector potential and velocity field. The \texttt{time} of the snapshot is given as a scalar. The \texttt{settings} and \texttt{grid} are optional and are stored in each snapshot by default.}

\graphfull{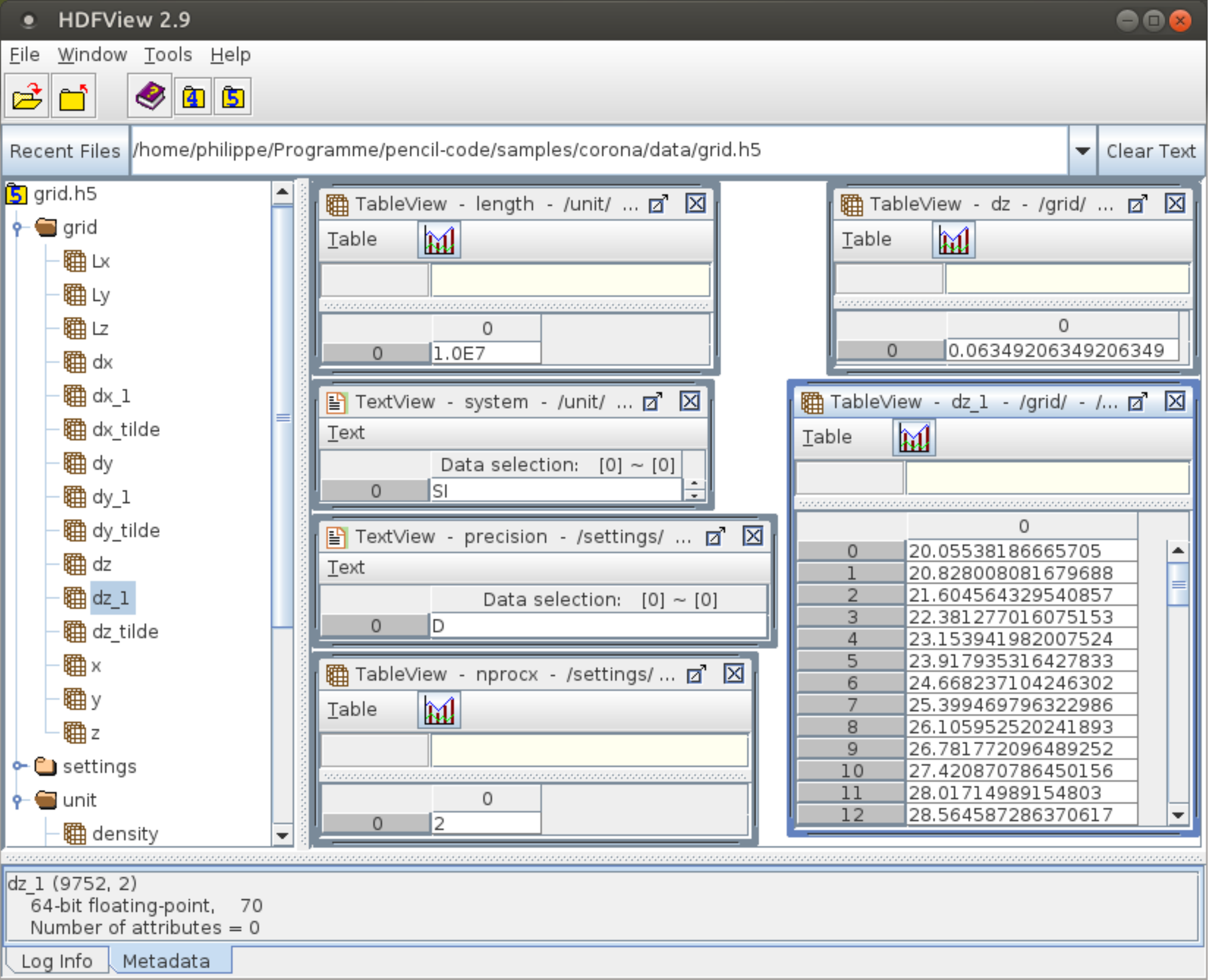}{gridfile}{Content of the file \texttt{`grid.h5'} from the \texttt{corona} sample, where we use a non-equidistant grid in the \texttt{z}-direction.
The unit length is \eqi{10^7} in \texttt{SI} units and we use double precision for floating-point numbers.
The array \texttt{dz\_1} contains the inverse grid spacing along the \texttt{z}-direction and the scalar \texttt{dz} gives the average grid spacing.}

The classical snapshot files like \texttt{var.h5} do now contain the grid positions, dimensions of the setup, as well as some basic simulation settings in a separate data structure.
This additional information can be suppressed to save storage space when a large number of small data snapshots is generated; set \texttt{lomit\_add\_data=T} within the section \texttt{run\_pars} inside the \texttt{run.in} configuration file.
The main simulation data is written out in components, like \texttt{ux}, \texttt{uy}, \texttt{uz}, \texttt{lnrho} or \texttt{rho}, etc.
Unused components are of course not written.
Inner ghost layers are cut for all quantities that are defined on grid cells.
The {\em{HDF5}} snapshots are hence always monolithic and are stored in the directory \texttt{data/allprocs}.

A typical example of the content of a \texttt{var.h5} snapshot can be visualized with the tool \texttt{`hdfview'}; see \fig{varfile}.
The datasets (like \texttt{ux}) are listed in groups (like \texttt{data} or \texttt{settings}) that can be opened and closed by a double mouseclick.
Some fundamental parameters of the simulation can be contained in snapshot files, e.g. \texttt{settings/precison} holds either an \texttt{`S'} for single or a \texttt{`D'} for double precision.
The time of the snapshot is stored as a separate scalar double-precision dataset named \texttt{time}.
Datasets can also be multi-dimensional arrays and the order of these arrays is in the canonical \texttt{Fortran} way, which means the first dimension of the array is along the \eqi{z}-direction.
This implies that one has to transpose multi-dimensional data arrays for post-processing with languages like \texttt{C} or \texttt{Julia}/\texttt{Python}, while the datasets are naturally aligned for languages like \texttt{Fortran} or \texttt{IDL}.

The so-called persistent data (that is usually associated per processor) is written to global arrays of the size (\texttt{nprocx},\texttt{nprocy},\texttt{nprocz}), where \texttt{nprocx} means the number of processors along the \texttt{x}-direction.
Particle data gets collected from all processors and is stored in global arrays together with the information of the mapping of particles to the processors.

In addition to the snapshots, the module \texttt{`io\_hdf5'} generates also a grid file \texttt{data/grid.h5} that contains the global grid data, as well as some fundamental parameters, like the size of the simulation domain, the number of grid points, grid distances, etc.; see \fig{gridfile}.

For inspecting {\em{HDF5}} files from a text-based command line, one can use the \texttt{`h5dump'} command.
The optional parameter \texttt{`-H'} allows to see the file structure (groups and datasets) without printing the actual data.

In \tab{timings} we compare the advantages and disadvantages of each IO method available in the {\em{Pencil Code}} for a setup with \eqi{1024 \times 1024 \times 256} grid cells distributed on \eqi{8 \times 16 \times 8 = 1024} processors.
One monolithic snapshot has a size of 17 GB and the timings were obtained on the {\em{JuRoPA}} supercomputer with the {\em{Lustre}} filesystem in a version from 2012.
Red entries in \tab{timings} highlight the disadvantages.
Please note these timings are hardly comparable with nowadays supercomputers, because the {\em{JuRoPA}} hardware and storage system are outdated and had beed decomissioned several years ago.

\section{Conclusions} \label{S:conclusion}

We use small-scale granulation and large-scale magnetic patches to drive a corona model by magnetic observations of an active region.
The combination of both horizontal-velocity fields extends the spectrum of driving velocities to cover a large range in amplitues and in spatial length scales; see \fig{LCT_histogram}.
An adaptice atmospheric boundary condition together with an adaptive chromospheric Newton cooling allows to formulate a flexible lower-boundary condition for the coronal part of the model.
The varying magnetic pressure around strong polarities or the granulation driver may increase the plasma density in the intergranular lanes.
This changes the density in the lower atmosphere and requires a matching response in from Newton cooling term.
The new Newton cooling scheme implemented in the {\em{Pencil Code}} allows for expanding and shrinking atmospheric columns at the lower boundary, so that the temperature stratification is now adaptive.

For the upper boundary we have developed several mechanisms to diffuse away perturbations in the density and temperature.
A new swamp-diffusion region can be used to relax non-force-free magnetic fields in order to reduce strong currents at the upper physical boundary, where ``open'' magnetic fields are desired but are in a force-free state.

\graphfull{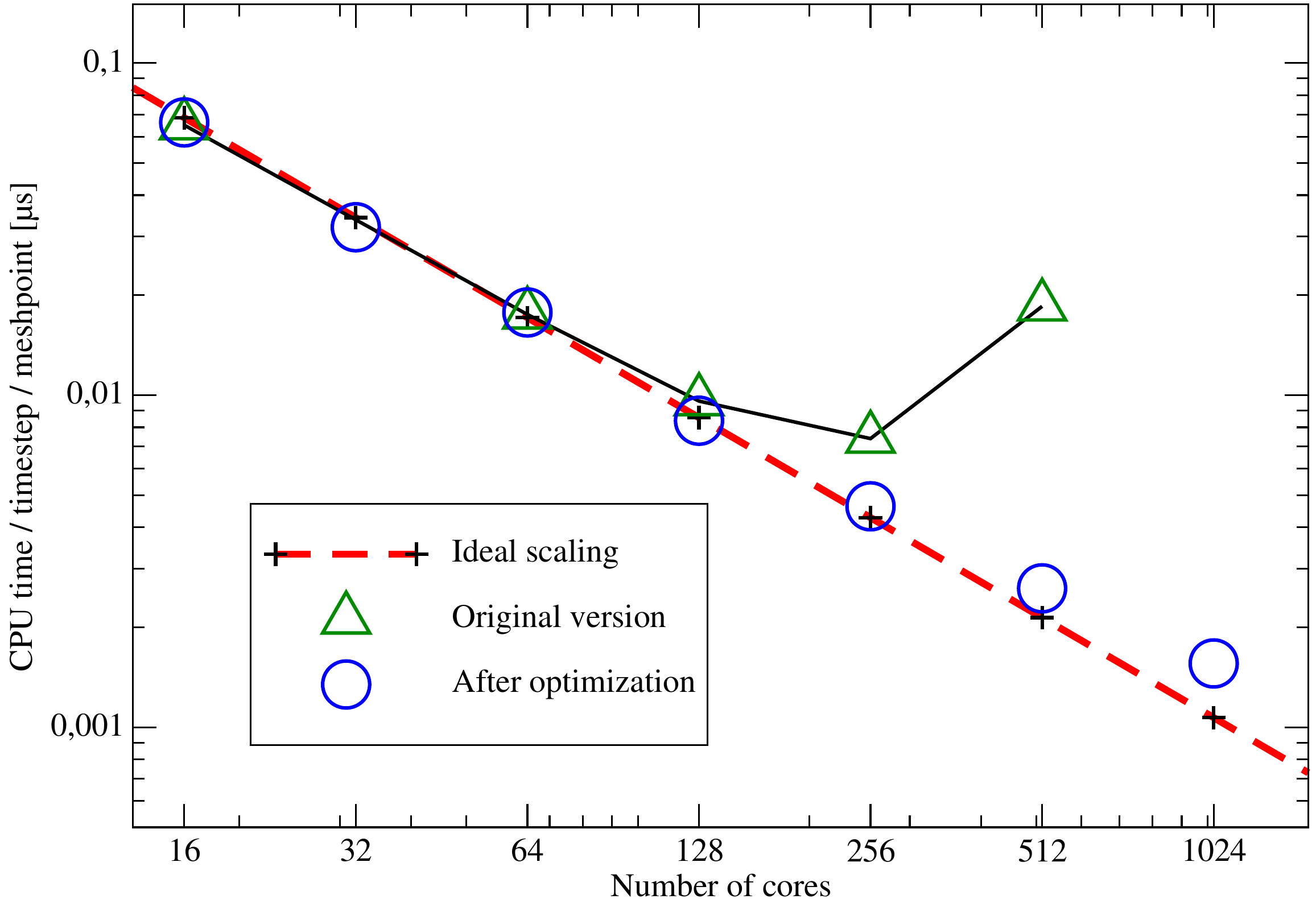}{hard-scaling}{Comparison of the hard scaling before (triangles) and after (circles) the implementation of a massive-parallel FFT in the {\em{Pencil Code}} versus the theoretical scaling limit (dashed red line).}

The IO strategy is crucial for large-scale models in nowadays science where large amounts of data are generated.
We find that only the modules \texttt{`io\_mpi2'} and \texttt{`io\_hdf5'} are optimal regarding storage requirements and scalability.
Of these two, only \texttt{`io\_hdf5'} offers a transparent, portable, and extendable data structure that allows to change the file formats without loosing backwards compatibility for the main code and for any data analysis scripts.
Furthermore, almost all modern data analysis languages provide at least reading routines for the {\em{HDF5}} file format.

Boundary conditions that make use of a FFT, like a potential-field extrapolation, benefit from massive-parallel FFT implementation, like the routine \texttt{`fft\_xy\_parallel'} provided in the \texttt{`fourier\_fftpack'} module.
The scalability of such boundary conditions is significantly improved as compared to the original \texttt{`fourier\_transform'} set of routines.

Altogether, the scalability of the {\em{Pencil Code}} could be improved substantially with massive-parallel methods that we implemented for the IO modules and for the FFT used by the potential-field boundary condition \texttt{\apos{pfe}}.
We show a comparison plot for the total runtime in \fig{hard-scaling} that we obtained with a constant global number of grid cells but for an increasing number of processors, which is usually called a `hard scaling' test.

The advantages of the {\em{HDF5}} file format become clear from the comparison in \tab{timings}.
We recommend to concentrate future developments of IO routines on the \texttt{`io\_hdf5'} module and to fade out support for older IO modules during the next years.

\section*{Acknowledgements}
\bibfont
The results of this research have been achieved using the PRACE Research Infrastructure resource \emph{Curie} based in France at TGCC, as well as \emph{JuRoPA} hosted by the J{\"u}lich Supercomputing Centre in Germany.
Hinode is a Japanese mission developed, launched, and operated by ISAS/JAXA, in partnership with NAOJ, NASA, and STFC (UK). Additional operational support is provided by ESA and NSC (Norway).

\disclosure{}
\ORCID{Philippe-A.~Bourdin}{0000-0002-6793-601X}

\bibliography{Literatur}
\bibliographystyle{gGAF}

\end{document}